\documentclass[aps,pra,english,twocolumn,notitlepage,showpacs,nofootinbib]{revtex4-1}
\usepackage{mathptmx}
\usepackage[T1]{fontenc}
\usepackage[latin9]{inputenc}
\usepackage{color}
\usepackage{babel}
\usepackage{textcomp}
\usepackage{bm}
\usepackage{graphicx}
\usepackage{subfigure}
\usepackage{amsmath}
\usepackage{amsthm}
\usepackage{amssymb}
\usepackage{epstopdf}
\usepackage[unicode=true,pdfusetitle,
bookmarks=true,bookmarksnumbered=false,bookmarksopen=false,
breaklinks=true,pdfborder={0 0 0},pdfborderstyle={},backref=false,colorlinks=true]{hyperref}
\usepackage{amsmath, amsfonts, amssymb, amsthm, mathrsfs, amssymb, braket, bbm, bm,graphicx,times,txfonts,color,subfigure}
\usepackage{hyperref}
\usepackage[table]{xcolor}

\hypersetup{
urlcolor=blue, citecolor=blue, linkcolor=blue}
\makeatletter
\theoremstyle{plain}
\theoremstyle{definition}
\newtheorem{Theorem}{Theorem}

\newtheorem{Lemma}{Lemma}

\newtheorem{Corollary}{Corollary}
\newtheorem{Conjecture}{Conjecture}
\makeatother
\begin{document}

\title{Bargmann invariants of Gaussian states}
\author{Jianwei Xu}
\email{xxujianwei@163.com}

\begin{abstract}
Given a set of ordered quantum states, described by density operators $%
\{\rho _{j}\}_{j=1}^{n}$, the Bargmann invariant of $\{\rho _{j}\}_{j=1}^{n}$
is defined as tr($\rho _{1}\rho _{2}...\rho _{n}$). Bargmann invariant
serves as a fundamental concept for quantum mechanics and has diverse
applications in quantum information science. Bosonic Gaussian states are a
class of quantum states on infinite-dimensional Hilbert space, widely used
in quantum optics and quantum information science. Bosonic Gaussian states
are conveniently and conventionally characterized by their means and
covariance matrices. In this work, we provide the expression of Bargmann
invariant tr($\rho _{1}\rho _{2}...\rho _{n}$) for any $m$-mode bosonic
Gaussian states $\{\rho _{j}\}_{j=1}^{n}$ in terms of the means and
covariance matrices of $\{\rho _{j}\}_{j=1}^{n}.$ We also use this
expression to explore the permissible values of Bargmann invariants for
bosonic Gaussian states.
\end{abstract}

\maketitle



\section{Introduction}

For a set of ordered quantum states, described by the density operators $%
\{\rho _{j}\}_{j=1}^{n}$ on a complex Hilbert space, the Bargmann invariant
\cite{Bargmann-1964-JMP}, or called the multivariate trace \cite{Wilde-2024-Quantum},  of $\{\rho _{j}\}_{j=1}^{n}$ is defined as tr($\rho _{1}\rho
_{2}...\rho _{n}$). A basic property of tr($\rho _{1}\rho _{2}...\rho _{n}$)
is that tr($\rho _{1}\rho _{2}...\rho _{n}$) is invariant under the
transformation $\{\rho _{j}\}_{j=1}^{n}\rightarrow \{U\rho _{j}U^{\dag
}\}_{j=1}^{n},$ here $U$ is a unitary operator, $U^{\dag }$ denotes the
adjoint of $U.$ This property implies that we can compute tr($\rho _{1}\rho
_{2}...\rho _{n}$) in arbitrary orthonormal basis of the Hilbert space.
Bargmann invariant has many applications in quantum information science and
links to many concepts, such as in geometric phases \cite{Simon-2001-PRA,Simon-2003-PRA,Simon-2003-JPA,Popov-2023-PRB}, photonic
indistinguishability \cite{Menssen-2017-PRL,Jones-2020-PRL}, quantum error mitigation
\cite{Fei-2023-PRA}, scalar spin chirality \cite{Galvao-2023-PRR}, Kirkwood-Dirac distribution \cite{Kirkwood-1933-PR,Dirac-1945-RMP,NYH-2021-JPA,Bievre-2024-NJP}, quantum coherence \cite{Brunner-2021-PRL,li-2025-arxiv}, quantum imaginarity \cite{Miyazaki-2022-Quantum,Fernandes-2024-PRL,li-2025-arxiv}, and weak values \cite{Wagner-2023-PRA}. Bargmann invariants can be measured using different experimental methods \cite{Halpern-2018-PRA,Pont-2022-PRX,Oszmaniec-2024-NJP,Wagner-2024-QST,Wilde-2024-Quantum,Galvao-2025-arxiv}.

(Bosonic) Gaussian states are a kind of infinite-dimensional quantum states (some recent reviews see for examples Refs. \cite{Braunstein-2005-RMP,Wang-2007-PR,Ferraro-2005-arxiv,Olivares-2012-EPJ,Weedbrook-2012-RMP,Adesso-2014-OSID,Serafini-2023-book}),
which are widely used in quantum optics and quantum information science.
Thermal states, Glauber coherent states, and squeezed states are special
cases of Gaussian states, which can be explicitly expressed in Fock basis. However, general Gaussian states are not easy to be expressed in Fock basis (some explorations about how to express general Gaussian states in Fock basis are reported in Refs. \cite{Xu-2016-PRA,Quesada-2019-PRA,Quesada-2024-SP}), but is conveniently and conventionally
expressed by its mean and covariance matrix. Therefore, in most cases, we
hope that a quantity or a property of Gaussian states can be expressed by the means and
covariance matrices of Gaussian states although it is often a hard problem, such as quantum entanglement \cite{Simon-2000-PRL,Duan-2000-PRL,Cirac-2001-PRL,Werner-2002-PRA,Adesso-2004-PRA,Marian-2008-PRA,Chen-2023-PRA}, quantum discord \cite{Adesso-2010-PRL,Paris-2010-PRL,Lloyd-2014-PRL}, quantum coherence
\cite{Xu-2016-PRA,Paris-2017-PRA,Du-2022-PRA,Du-2023-PRA}, quantum correlations \cite{Qi-2022-PRA}, and quantum fidelity \cite{Marian-2012-PRA,Banchi-2015-PRL}.
The Bargmann invariant tr($\rho_{1}\rho_{2}$) for any two $m$-mode Gaussian states $\{\rho _{j}\}_{j=1}^{2}$ has been well characterized in terms of their means and covariance matrices. However, the generalization of this expression to $n$-order Bargmann invariant tr($\rho _{1}\rho _{2}...\rho _{n}$) for arbitrary $m$-mode Gaussian states $\{\rho _{j}\}_{j=1}^{n}$ with $n>2$ remains an open problem.

In this work, we provide the explicit expression of Bargmann invariant tr($%
\rho _{1}\rho _{2}...\rho _{n}$) for any set of $m$-mode Gaussian states $%
\{\rho _{j}\}_{j=1}^{n}$ in terms of the means and covariance matrices of $%
\{\rho _{j}\}_{j=1}^{n}.$ The rest of this paper is organized as follows. In section
\hyperlink{section II}{II}, we review the definition of Gaussian states and some results of
Bargmann invariants. In section \hyperlink{section III}{III}, we give the expression of Bargmann
invariant tr($\rho _{1}\rho _{2}...\rho _{n}$) for any set of $m$-mode
Gaussian states $\{\rho _{j}\}_{j=1}^{n}$ in terms of the means and
covariance matrices of $\{\rho _{j}\}_{j=1}^{n}.$  In section \hyperlink{section IV}{IV}, we give some examples to
illustrate the application of the expression of Bargmann invariant tr($\rho
_{1}\rho _{2}...\rho _{n}$) and explore the permissible values of one-mode pure Gaussian states. Section \hyperlink{section V}{V} is a short summary. For
clarity of structure, we postpone some proofs to \hyperlink{Appendix-A}{Appendices}.

\hypertarget{section II}{}
\section{Gaussian states and Bargmann invariants}

\subsection{background of Gaussian states}

We review the definition and some basic facts about (bosonic) Gaussian
states and introduce the notation we will use.

Suppose $\{|l\rangle \}_{l=0}^{\infty }$ is an orthonormal basis with $l\in
\{0,1,2,3,...\},$ $\{|l\rangle \}_{l=0}^{\infty }$ spans the complex Hilbert
space $\mathcal{H}$ over the complex field $\mathbb{C}.$ $\{|l\rangle \}_{l=0}^{\infty }$ is called the one-mode Fock basis. The $m$-fold tensor product of $\{|l\rangle \}_{l=0}^{\infty }$, denoted by $%
\{|l\rangle \}_{l}^{\otimes m},$ is called the $m$-fold Fock basis, $%
\{|l\rangle \}_{l}^{\otimes m}$ then spans the complex Hilbert space $%
\mathcal{H}^{\otimes m}=\otimes _{k=1}^{m}\mathcal{H}_{k}$ with each $%
\mathcal{H}_{k}=\mathcal{H}$ over $\mathbb{C}.$ On each $\mathcal{H}_{k},$ annihilation operator $\widehat{a}_{l}$ and
creation operator $\widehat{a}_{l}^{\dagger }$ are defined as
\begin{eqnarray}
\widehat{a}_{k}|0\rangle &=&0,\ \ \widehat{a}_{k}|l\rangle =\sqrt{l}%
|l-1\rangle \text{ for }l\geq 1;  \notag  \label{eq2-1} \\
\widehat{a}_{k}^{\dagger }|l\rangle &=&\sqrt{l+1}|l+1\rangle \text{ for }%
l\geq 0.   \notag \label{eq2-2}
\end{eqnarray}%
We write $\{\widehat{a}_{k},\widehat{a}_{k}^{\dagger }\}_{k=1}^{m}$ as a
vector
\begin{eqnarray}
\widehat{A} &=&(\widehat{a}_{1},\widehat{a}_{1}^{\dagger },\widehat{a}_{2},%
\widehat{a}_{2}^{\dagger },...,\widehat{a}_{m},\widehat{a}_{m}^{\dagger
})^{T}  \notag \\
&=&(\widehat{A}_{1},\widehat{A}_{2},\widehat{A}_{3},\widehat{A}_{4},...,%
\widehat{A}_{2m-1},\widehat{A}_{2m})^{T},   \notag \label{eq2-3}
\end{eqnarray}%
where $T$ denotes the transposition.

From operators $\{\widehat{a}_{k},\widehat{a}_{k}^{\dagger }\}_{k=1}^{m}$,
the quadrature operators $\{\widehat{q}_{k},\widehat{p}_{k}\}_{k=1}^{m}$ are defined as
\begin{equation}
\widehat{q}_{k}=\widehat{a}_{k}+\widehat{a}_{k}^{\dagger },\ \ \widehat{p}%
_{k}=-i(\widehat{a}_{k}-\widehat{a}_{k}^{\dagger }),   \notag \label{eq2-4}
\end{equation}%
where $i=\sqrt{-1}.$ We also write $\{\widehat{q}_{k},\widehat{p}%
_{k}\}_{k=1}^{m}$ as a vector
\begin{eqnarray}
\widehat{X} &=&(\widehat{q}_{1},\widehat{p}_{1},\widehat{q}_{2},\widehat{p}%
_{2},...,\widehat{q}_{m},\widehat{p}_{m})^{T}  \notag \\
&=&(\widehat{X}_{1},\widehat{X}_{2},\widehat{X}_{3},\widehat{X}_{4},...,%
\widehat{X}_{2m-1},\widehat{X}_{2m})^{T}.   \notag \label{eq2-5}
\end{eqnarray}
With these definitions, we have the canonical commutation relations
\begin{eqnarray}
\lbrack \widehat{A}_{j},\widehat{A}_{k}]=\Omega _{jk},  \ \
\lbrack \widehat{X}_{j},\widehat{X}_{k}]=2i\Omega _{jk},   \notag \label{eq2-6}
\end{eqnarray}%
where $[\widehat{A}_{j},\widehat{A}_{k}]=\widehat{A}_{j}\widehat{A}_{k}-%
\widehat{A}_{k}\widehat{A}_{j}$ is the commutator of $\widehat{A}_{j}$ and $%
\widehat{A}_{k},$ $\Omega _{jk}$ is the entry of the $2m\times 2m$ matrix $%
\Omega $ with
\begin{equation}
\Omega =\oplus _{k=1}^{m}\omega ,\ \ \omega =\left(
\begin{array}{cc}
0 & 1 \\
-1 & 0%
\end{array}%
\right).   \label{eq2-7}
\end{equation}

A quantum state $\rho $ on $\mathcal{H}^{\otimes m}$ can be characterized by
its characteristic function
\begin{equation}
\chi (\rho ,\xi )=\text{tr}[\rho D(\xi )],   \notag \label{eq2-8}
\end{equation}%
where $D(\xi )$ is the displacement operator
\begin{eqnarray}
D(\xi ) &=&\exp (i\widehat{X}^{T}\Omega \xi ),  \label{eq2-9} \\
\xi  &=&(\xi _{1},\xi _{2},...,\xi _{2m})^{T}\in\mathbb{R}
^{2m}.  \label{eq2-10}
\end{eqnarray}

When $m=1,$ $\xi =(\xi _{1},\xi _{2})^{T}\in
\mathbb{R}^{2},$ $D(\xi )$ is often written as $D(\xi )=D(\alpha )=\exp (\alpha
\widehat{a}^{\dagger }-\alpha ^{\ast }\widehat{a}),$ here $\alpha =\xi
_{1}+i\xi _{2}\in
\mathbb{C},$ $a^{\dagger }=\widehat{a}_{1}^{\dagger },$ $\widehat{a}=\widehat{a}_{1},$
$\ast $ denotes the complex conjugation.

The mean of $\rho $ is defined as
\begin{equation}
\overline{X}=\text{tr}(\rho \widehat{X})=(\overline{X}_{1},\overline{X}%
_{2},...,\overline{X}_{2m})^{T};   \notag \label{eq2-11}
\end{equation}%
the covariance matrix $V$ is defined by its entries
\begin{equation}
V_{jk}=\frac{1}{2}\text{tr}(\rho \{\Delta \widehat{X}_{j},\Delta \widehat{X}%
_{k}\})  \label{eq2-12}
\end{equation}%
with $\Delta \widehat{X}_{j}=\widehat{X}_{j}-\overline{X}_{k}$, and $%
\{\Delta \widehat{X}_{j},\Delta \widehat{X}_{k}\}=\Delta \widehat{X}%
_{j}\Delta \widehat{X}_{k}+\Delta \widehat{X}_{k}\Delta \widehat{X}_{j}$ is
the anticommutator of $\Delta \widehat{X}_{j}$ and $\Delta \widehat{X}_{k}.$
The covariance matrix $V=V^{T}$ is a $2m\times 2m$ real and symmetric matrix
which satisfies the uncertainty principle \cite{Simon-1994-PRA}
\begin{equation}
V+i\Omega \succeq 0,   \label{eq2-13}
\end{equation}%
that is, $V+i\Omega $ is positive semidefinite. Note that $V+i\Omega \succeq
0$ implies $V\succ 0$ meaning that $V$ is positive definite.

With these preparations, we turn to the definition of Gaussian states. A
quantum state $\rho $ on $\mathcal{H}^{\otimes m}$ is called an $m$-mode
Gaussian state if its characteristic function has the Gaussian form
\begin{equation}
\chi (\rho ,\xi )=\exp \left[ -\frac{1}{2}\xi ^{T}(\Omega V\Omega ^{T})\xi
-i(\Omega \overline{X})^{T}\xi \right] ,  \label{eq2-14}
\end{equation}%
where $\overline{X}$ is the mean of $\rho $ and $V$ is the covariance matrix
of $\rho .$ The Gaussian state $\rho $ is determined by its characteristic
function $\chi (\rho ,\xi )$ via the inverse relation \cite{Serafini-2023-book}
\begin{equation}
\rho =\int \frac{d^{2m}\xi }{\pi ^{m}}\chi (\rho ,\xi )D(-\xi ),  \label{eq2-15}
\end{equation}%
where $\int =\int_{-\infty }^{\infty }.$ $\overline{X}$ and $V$ with Eq. (\ref{eq2-13}) completely determine the Gaussian state $\rho $ \cite{Simon-1994-PRA}, thus we write $\rho $ as $\rho (\overline{X},V).$

The complex conjugate in Fock basis of the $m$-mode Gaussian state $\rho (\overline{X},V)$ is still a Gaussian state $\rho (\overline{X}^{\prime},V^{\prime })$ with \cite{Xu-2023-PRA}
\begin{equation}
\overline{X}^{\prime }=(\oplus _{k=1}^{m}Z)\overline{X}, \ \
V^{\prime}=(\oplus _{k=1}^{m}Z)V(\oplus _{k=1}^{m}Z),   \label{eq2-16}
\end{equation}%
where $Z=\left(
\begin{array}{cc}
1 & 0 \\
0 & -1%
\end{array}%
\right) $ is one of the Pauli matrices.

\subsection{Bargmann invariants}

For quantum states $\{\rho _{j}\}_{j=1}^{n}$ on a $d$-dimensional complex
Hilbert space $\mathbb{C}^{d},$ the Bargmann invariant of $\{\rho _{j}\}_{j=1}^{n}$ is defined as tr$%
(\rho _{1}\rho _{2}...\rho _{n}).$ The set $\mathcal{B}_{n,d}$ is defined by
\begin{equation}
\mathcal{B}_{n,d}=\{\text{tr}(\rho _{1}\rho _{2}...\rho _{n}):\text{all
states }\{\rho _{j}\}_{j=1}^{n}\text{ on }%
\mathbb{C}^{d}\},     \notag \label{eq2-17}
\end{equation}
i.e., $\mathcal{B}_{n,d}$ is the set of all permissible values of $n$-order Bargmann
invariants on $\mathbb{C}^{d}$. In practice, we are mainly interested in the set
\begin{equation}
\mathcal{B}_{n}=\cup _{d=1}^{\infty }\mathcal{B}_{n,d}.    \notag \label{eq2-18}
\end{equation}%
Note that, when the dimensions $d_{1}<d_{2},$ we can always regard $%
\mathbb{C}^{d_{1}}$ as a subspace of $\mathbb{C}^{d_{2}},$ then it follows that
\begin{equation}
\mathcal{B}_{n,d_{1}}\subseteq \mathcal{B}_{n,d_{2}}.     \notag \label{eq2-19}
\end{equation}

Since $\{\rho _{j}^{\ast }\}_{j=1}^{n}$ are still density matrices if $\{\rho
_{j}\}_{j=1}^{n}$ are density matrices in a fixed orthonormal basis of the
complex Hilbert space and
\begin{equation}
\text{tr}(\rho _{1}^{\ast }\rho _{2}^{\ast }...\rho _{n}^{\ast })=[\text{tr}%
(\rho _{1}\rho _{2}...\rho _{n})]^{\ast },   \label{eq2-20}
\end{equation}%
then
\begin{equation}
\mathcal{B}_{n}=\mathcal{B}_{n}^{\ast },   \label{eq2-21}
\end{equation}%
where $\mathcal{B}_{n}^{\ast }=\{z^{\ast }:z\in \mathcal{B}_{n}\}.$

Clearly,
\begin{equation}
\mathcal{B}_{1}=\{1\}, \ \ \mathcal{B}_{2}=[0,1].    \label{eq2-22}
\end{equation}%
Rigorously determining the set $\mathcal{B}_{n}$ for $n\geq 3$ is not an
easy task. With the efforts of Refs. \cite{Fernandes-2024-PRL,Oszmaniec-2024-NJP,Li-2025-PRA,Zhang-2025-PRA,Xu-2025-arxiv,Pratapsi-2025-arXiv}, the set $\mathcal{B}_{n}$
for $n\geq 3$ was finally completely determined as \cite{Xu-2025-arxiv,Pratapsi-2025-arXiv}
\begin{equation}
\mathcal{B}_{n}=\left\{te^{i\theta }\cos ^{n}\frac{\pi }{n}\sec ^{n}\frac{\pi
-\theta }{n}:t\in \lbrack 0,1],\theta \in \lbrack 0,2\pi )\right\}.   \label{eq2-23}
\end{equation}
The boundary of $\mathcal{B}_{n}$ is
\begin{equation}
\partial\mathcal{B}_{n}=\left\{e^{i\theta }\cos ^{n}\frac{\pi }{n}\sec ^{n}\frac{\pi
-\theta }{n}:\theta \in \lbrack 0,2\pi )\right\}.   \label{eq2-23-1}
\end{equation}

When we focus on the Gaussian states, we define the set
\begin{equation}
\mathcal{G}_{n,m}=\{\text{tr}(\rho _{1}\rho _{2}...\rho _{n}):\text{all }m%
\text{-mode Gaussian states }\{\rho _{j}\}_{j=1}^{n}\},   \label{eq2-24}
\end{equation}%
that is, $\mathcal{G}_{n,m}$ is the set of all permissible values of any set
of $m$-mode Gaussian states $\{\rho _{j}\}_{j=1}^{n}.$

Clearly,
\begin{equation}
\mathcal{G}_{n,m}\subseteq \mathcal{B}_{n}\subsetneqq \{z:z\in\mathbb{C},|z|\leq 1\},   \ \ \ \
\mathcal{G}_{1,m}=\mathcal{B}_{1}=\{1\},  \label{eq2-25}
\end{equation}
where $\{z:z\in\mathbb{C},|z|\leq 1\}$ represents the unit disk in the complex plane.

For the modes $m_{1}<m_{2},$ we have
\begin{equation}
\mathcal{G}_{n,m_{1}}\subseteq \mathcal{B}_{n,m_{2}}.   \label{eq2-26}
\end{equation}%
A proof of Eq. (\ref{eq2-26}) is as follows. Suppose $\{\rho _{j}\}_{j=1}^{n}$ are all $%
m_{1}$-mode Gaussian states. Note that the vacuum state $|0\rangle \langle 0|
$ is a one-mode Gaussian state, then $\{\rho _{j}\otimes (|0\rangle \langle
0|)^{\otimes (m_{2}-m_{1})}\}_{j=1}^{n}$ are all $m_{2}$-mode Gaussian
states, and
\begin{eqnarray}
&&\text{tr}(\rho _{1}\rho _{2}...\rho _{n}) \notag \\
&=&\text{tr}\left\{ [(\rho _{1}\otimes (|0\rangle \langle 0|)^{\otimes
(m_{2}-m_{1})}]...[(\rho _{n}\otimes (|0\rangle \langle 0|)^{\otimes
(m_{2}-m_{1})}]\right\}. \notag \ \ \
\end{eqnarray}%
This implies Eq. (\ref{eq2-26}). Similar to Eq. (\ref{eq2-21}), since $\{\rho _{j}^{\ast
}\}_{j=1}^{n}$ are still Gaussian states if $\{\rho _{j}\}_{j=1}^{n}$ are
Gaussian states \cite{Xu-2023-PRA} and Eq. (\ref{eq2-20}) holds, then
\begin{equation}
\mathcal{G}_{n,m}=\mathcal{G}_{n,m}^{\ast }.   \notag \label{eq2-27}
\end{equation}

If $z\in \mathcal{B}_{n}$ ($z\in \mathcal{G}_{n,m}$), we say $z$ is a
permissible value or realizable value of $\mathcal{B}_{n}$ ($\mathcal{G}%
_{n,m}$).

\hypertarget{section III}{}
\section{Bargmann invariants of Gaussian states}

In this section, we derive the expression of Bargmann invariants of Gaussian
states. Theorem 1 below is the main result of this work.
\begin{Theorem} \label{Theorem-1}
For $n\geq 2,$ the Bargmann invariant tr$(\rho _{1}\rho
_{2}...\rho _{n})$ of the ordered $m$-mode Gaussian states $\{\rho
_{j}\}_{j=1}^{n}$ can be expressed by
\begin{eqnarray}
&&\text{tr}(\rho _{1}\rho _{2}...\rho _{n}) =\frac{2^{m(n-1)}}{\sqrt{\det M}}%
\exp \left( -\frac{1}{2}\Lambda ^{T}M^{-1}\Lambda \right) ,   \label{eq3-1} \\
&&M=\left(
\begin{array}{ccccc}
V^{(n)}+V^{(1)} & V^{(n)}+i\Omega  & V^{(n)}+i\Omega  & ... &
V^{(n)}+i\Omega  \\
V^{(n)}-i\Omega  & V^{(n)}+V^{(2)} & V^{(n)}+i\Omega  & ... &
V^{(n)}+i\Omega  \\
V^{(n)}-i\Omega  & V^{(n)}-i\Omega  & V^{(n)}+V^{(3)} & ... &
V^{(n)}+i\Omega  \\
... & ... & ... & ... & ... \\
V^{(n)}-i\Omega  & V^{(n)}-i\Omega  & V^{(n)}-i\Omega  & ... &
V^{(n)}+V^{(n-1)}%
\end{array}%
\right) , \notag \\   \label{eq3-2}  \\
&&\Lambda=\left(
\begin{array}{c}
\overline{X}^{(1)}-\overline{X}^{(n)} \\
\overline{X}^{(2)}-\overline{X}^{(n)} \\
... \\
\overline{X}^{(n-1)}-\overline{X}^{(n)}%
\end{array}%
\right),   \label{eq3-3}
\end{eqnarray}%
where $\overline{X}^{(j)}$ and $V^{(j)}$ are the mean and covariance matrix
of $\rho _{j},$ $M^{-1}$ is the inverse of $M$.
\end{Theorem}

Notice that the square root function has two branches. That is, if we write
the complex number $z$ as $z=re^{i\theta }$ with $r>0$ and $\theta \in \lbrack
0,2\pi ),$ then the square root function has two branches $\sqrt{z}=\sqrt{r}%
\exp (i\frac{\theta }{2})$ and $-\sqrt{z}=-\sqrt{r}\exp (i\frac{\theta }{2}).
$ $\sqrt{z}=\sqrt{r}\exp (i\frac{\theta }{2})$ is called the principal
branch of the square root function. In Eq. (\ref{eq3-1}), $\sqrt{\det M}$ takes the
principal branch of the square root function of $\det M$.

Notice also that since the Bargmann invariant $\text{tr}(\rho _{1}\rho _{2}...\rho _{n})$ has the cyclic invariance, that is, $\text{tr}(\rho _{1}\rho _{2}...\rho _{n})=\text{tr}(\rho _{2}...\rho _{n}\rho _{1})=...,$   then the right-hand side of Eq. (\ref{eq3-1}) also has the cyclic invariance.

We give a proof of Theorem \ref{Theorem-1} in Appendix \hyperlink{Appendix-A}{A.}

Note that when $n=2,$ Theorem \ref{Theorem-1} reduces to the well-known overlap formula
\cite{Serafini-2023-book}
\begin{eqnarray}
&&\text{tr}(\rho _{1}\rho _{2})=\frac{2^{m}}{\sqrt{\det (V^{(2)}+V^{(1)})}} \notag \\
&&\ \ \ \ \cdot\exp \left[ -\frac{1}{2}(\overline{X}^{(1)}-\overline{X}%
^{(2)})^{T}(V^{(2)}+V^{(1)})^{-1}(\overline{X}^{(1)}-\overline{X}^{(2)})%
\right]. \ \ \ \   \label{eq3-4}
\end{eqnarray}

When all $m$-mode Gaussian states $\{\rho _{j}\}_{j=1}^{n}$ are the same
state $\rho _{j}=\rho ,$ Theorem \ref{Theorem-1} becomes Corollary 1 below.
\begin{Corollary} \label{Corollary-1}
For $n\geq 2$ and the $m$-mode Gaussian state $\rho (\overline{X},V),$ it holds that
\begin{eqnarray}
\text{tr}(\rho ^{n}) &=&\frac{2^{m(n-1)}}{\sqrt{\det M_{V} }},  \label{eq3-5}  \\
M_{V} &=&\left(
\begin{array}{ccccc}
2V & V+i\Omega & V+i\Omega & ... & V+i\Omega \\
V-i\Omega & 2V & V+i\Omega & ... & V+i\Omega \\
V-i\Omega & V-i\Omega & 2V & ... & V+i\Omega \\
... & ... & ... & ... & ... \\
V-i\Omega & V-i\Omega & V-i\Omega & ... & 2V%
\end{array}%
\right).   \label{eq3-6}
\end{eqnarray}
\end{Corollary}

There is another way to compute tr$(\rho ^{n})$ by the thermal decomposition
of $\rho .$ Any $m$-mode Gaussian state $\rho (\overline{X},V)$ allows a
thermal decomposition as \cite{Serafini-2023-book}
\begin{eqnarray}
\rho (\overline{X},V) &=&D(\overline{X})U_{S}[\rho (0,V^{\oplus
})]U_{S}^{\dagger }D(\overline{X})^{\dagger },   \notag \label{eq3-7}  \\
U_{S}[\rho (0,V^{\oplus })]U_{S}^{\dagger } &=&\rho (0,V),   \notag \label{eq3-8} \\
V &=&SV^{\oplus }S^{T},  \label{eq3-9} \\
V^{\oplus } &=&\oplus _{k=1}^{m}\nu _{k}I_{2},\nu _{k}\geq 1,  \label{eq3-10} \\
S\Omega S^{T} &=&\Omega ,  \label{eq3-11}
\end{eqnarray}%
where $D(\overline{X})$ is the displacement operator which is a unitary
operator, $U_{S}$ is a unitary operator associated with a real symplectic matrix $%
S,$ $I_{2}$ is the identity matrix of size $2.$ A real matrix $S$ is called
a symplectic matrix if $S$ satisfies $S\Omega S^{T}=\Omega .$ A symplectic matrix $S$ must satisfy det$S=1$ \cite{Rump-2017-LAA}. $\{\nu
_{k}\}_{k=1}^{m}$ is called the symplectic eigenvalues of $V.$ $\rho
(0,V^{\oplus })$ is the $m$-mode thermal state, expressed in Fock basis as
\begin{eqnarray}
\rho (0,V^{\oplus }) &=&\oplus _{k=1}^{m}\rho _{\text{th}}\left( \overline{%
n_{k}}\right) ,   \notag \label{eq3-12} \\
\rho _{\text{th}}\left( \overline{n_{k}}\right)  &=&\sum_{l=0}^{\infty }%
\frac{\overline{n_{k}}^{l}}{(\overline{n_{k}}+1)^{l+1}}|l\rangle \langle l|,   \notag \label{eq3-13}  \\
\overline{n_{k}} &=&\frac{\nu _{k}-1}{2},    \notag \label{eq3-14}
\end{eqnarray}%
where $\overline{n_{k}}\geq 0$ is called the mean particle number (for
example, mean photon number in an optical field) of $\rho _{\text{th}%
}\left( \overline{n_{k}}\right) $ because of the fact $\overline{n_{k}}=$tr$%
\left[ \widehat{a}_{k}^{\dagger }\widehat{a}_{k}\rho _{\text{th}}\left(
\overline{n_{k}}\right) \right] .$ $\rho (\overline{X},V)$ is pure if and only if all symplectic eigenvalues $\nu _{k}=1,$ or equivalently, det$V=1.$
Employing the thermal decomposition, we can compute tr$(\rho ^{n})$ as
\begin{eqnarray}
\text{tr}(\rho ^{n}) &=&\text{tr}\left[ \oplus _{k=1}^{m}(\rho _{\text{th}%
}\left( \overline{n_{k}}\right) )^{n}\right]  \notag \\
&=&\prod_{k=1}^{m}\frac{1}{(\overline{n_{k}}+1)^{n}-\overline{n_{k}}^{n}}   \notag  \\
&=&\prod_{k=1}^{m}\frac{2^{n}}{(\nu _{k}+1)^{n}-(\nu _{k}-1)^{n}}.  \label{eq3-15}
\end{eqnarray}

Comparing above two expressions of tr$(\rho ^{n})$ in Corollary \ref{Corollary-1} and in Eq. (\ref{eq3-15}),
we see that Eq. (\ref{eq3-15}) depends on the symplectic eigenvalues $\{\nu
_{k}\}_{k=1}^{m}$ of $V.$ However, computing the symplectic eigenvalues $%
\{\nu _{k}\}_{k=1}^{m}$ of $V$ is not easy for general $V.$ In
contrast, the expression tr$(\rho ^{n})$ in Corollary \ref{Corollary-1} only depends on $\sqrt{%
\det M},$ and $\sqrt{\det M}$ can be directly computed for given $V.$ In conclusion, the
expression tr$(\rho ^{n})$ in Corollary \ref{Corollary-1} is more advantageous than the
expression tr$(\rho ^{n})$ in Eq. (\ref{eq3-15}) for general Gaussian states from the viewpoint of computation. To
demonstrate the application of Corollary \ref{Corollary-1} and reveal the intrinsic relationship
between these two expressions, in
Appendix \hyperlink{Appendix-B}{B}, we give a direct proof that these two expressions of tr$(\rho ^{n})$ are indeed equal.

\hypertarget{section IV}{}
\section{Permissible values of Bargmann invariants of one-mode pure
Gaussian states}

In this section, we provide some examples of one-mode pure states to
demonstrate the application of Theorem \ref{Theorem-1} and explore the region of $\mathcal{%
G}_{n,1}.$

Any one-mode pure Gaussian state can be expressed as a squeezed coherent
state
\begin{equation}
|\zeta ,\alpha \rangle =D(\alpha )\exp \left[\frac{1}{2}(\zeta ^{\ast }\widehat{a}%
^{2}-\zeta \widehat{a}^{\dagger 2})\right]|0\rangle ,   \notag \label{eq4-1}
\end{equation}%
where $\zeta ,\alpha $ are complex numbers, $\exp \left[\frac{1}{2}(\zeta ^{\ast }%
\widehat{a}^{2}-\zeta \widehat{a}^{\dagger 2})\right]$ is the squeezing operator.
We write $\zeta ,\alpha $ in the polar form as $\zeta =|\zeta |e^{i\phi },$ $%
\alpha =|\alpha |e^{i\varphi }.$ The mean $\overline{X}$ and covariance
matrix $V$ of $|\zeta ,\alpha \rangle $ are
\begin{eqnarray}
&& \ \ \ \ \  \overline{X}=2(\text{Re}\alpha ,\text{Im}\alpha )^{T},  \label{eq4-2}  \\
&&\begin{cases}
V_{11}=\cosh (2|\zeta |)+\cos \phi \sinh (2|\zeta |) \\
V_{12}=V_{21}=\sin \phi \sinh (2|\zeta |) \\
V_{22}=\cosh (2|\zeta |)-\cos \phi \sinh (2|\zeta |).%
\end{cases}    \label{eq4-3}
\end{eqnarray}

When $\zeta =0,$ $|\zeta ,\alpha \rangle $ becomes the Glauber coherent state
\begin{equation}
|\alpha \rangle =D(\alpha )|0\rangle =e^{-\frac{|\alpha |^{2}}{2}%
}\sum_{l=0}^{\infty }\frac{\alpha ^{l}}{\sqrt{l!}}|l\rangle .    \notag \label{eq4-4}
\end{equation}%
For two Glauber coherent states $|\alpha _{1}\rangle $ and $|\alpha
_{2}\rangle ,$ one has \cite{Barnett-2002-book}
\begin{equation}
\langle \alpha _{1}|\alpha _{2}\rangle =\exp \left[ \alpha _{1}^{\ast
}\alpha _{2}-\frac{1}{2}(|\alpha _{1}|^{2}+|\alpha _{2}|^{2})\right] .   \label{eq4-5}
\end{equation}

When $\alpha =0,$ $|\zeta ,\alpha \rangle $ becomes the squeezed state
\begin{eqnarray}
|\zeta \rangle  &=&\exp \left[\frac{1}{2}(\zeta ^{\ast }\widehat{a}^{2}-\zeta
\widehat{a}^{\dagger 2})\right]|0\rangle  \notag \\
&=&\frac{1}{\sqrt{\cosh |\zeta |}}\sum_{l=0}^{\infty }(-e^{i\theta }\tanh
|\zeta |)^{l}\frac{\sqrt{(2l)!}}{2^{l}l!}|2l\rangle .\ \    \notag \label{eq4-6}
\end{eqnarray}%
For two squeezed states $|\zeta _{1}\rangle ,$ $|\zeta _{2}\rangle ,$ with $%
\zeta _{1}=|\zeta _{1}|e^{i\phi _{1}}$ and $\zeta _{2}=|\zeta _{2}|e^{i\phi
_{2}}$ their polar forms, it holds that \cite{Barnett-2002-book}
\begin{equation}
\langle \zeta _{1}|\zeta _{2}\rangle =\frac{1}{\sqrt{\cosh |\zeta _{1}|\cosh
|\zeta _{2}|-e^{i(\phi _{2}-\phi _{1})}\sinh |\zeta _{1}|\sinh |\zeta _{2}|}},   \label{eq4-7}
\end{equation}
where the square root takes the principal branch.

The expression of general $|\zeta ,\alpha \rangle $ in Fock basis is rather
complex \cite{Knight-2004-book}. Employing Eqs. (\ref{eq2-16},\ref{eq4-2},\ref{eq4-3}), one finds
\begin{equation}
|\zeta ,\alpha \rangle ^{\ast }=|\zeta ^{\ast },\alpha ^{\ast }\rangle , \ \
|\alpha \rangle ^{\ast }=|\alpha ^{\ast }\rangle , \ \
|\zeta\rangle ^{\ast }=|\zeta ^{\ast }\rangle .     \notag \label{eq4-8}
\end{equation}

Below we consider some special sets of one-mode squeezed coherent states to
explore the region of $\mathcal{G}_{n,1}.$

\textbf{Example 1.} Consider the ordered set of one-mode pure Gaussian states $%
\{|\Phi _{j}\rangle \}_{j=1}^{n},$
\begin{eqnarray}
|\Phi _{j}\rangle=\left\vert \zeta ,\alpha _{j}\right\rangle ,  \ \
\alpha _{j} =|\alpha |\exp \left( i\frac{2\pi }{n}j\right) .    \label{eq4-9}
\end{eqnarray}

The mean of $|\Phi _{j}\rangle $ is $\overline{X}^{(j)}=2|\alpha |\left(\cos
\frac{2\pi j}{n},\sin \frac{2\pi j}{n}\right)^{T},$ the covariance matrix $%
V^{(j)}=V$ with $V$ expressed in Eq. (\ref{eq4-3}). Eq. (\ref{eq3-3}) yields
\begin{eqnarray}
\Lambda =2|\alpha |\left(
\begin{array}{c}
(\cos \frac{2\pi }{n}-1,\sin \frac{2\pi }{n})^{T} \\
(\cos \frac{2\pi \cdot 2}{n}-1,\sin \frac{2\pi \cdot 2}{n})^{T} \\
... \\
(\cos \frac{2\pi (n-1)}{n}-1,\sin \frac{2\pi (n-1)}{n})^{T}%
\end{array}%
\right).   \label{eq4-10}
\end{eqnarray}

From Theorem \ref{Theorem-1} we see that, to compute tr$\left( |\Phi _{1}\rangle \left\langle \Phi
_{1}\right\vert \Phi _{2}\rangle \langle \Phi _{2}|...\Phi _{n}\rangle
\left\langle \Phi _{n}\right\vert \right) ,$ we need to compute $\det M$ and
$M^{-1}.$ We denote the matrix $M$ defined in Eq. (\ref{eq3-2}) with $V^{(j)}=V$
expressed in Eq. (\ref{eq4-3}) by $M_{Vp},$ that is,
\begin{equation}
M_{Vp}=\left(
\begin{array}{ccccc}
2V & V+i\omega & V+i\omega & ... & V+i\omega \\
V-i\omega & 2V & V+i\omega & ... & V+i\omega \\
V-i\omega & V-i\omega & 2V & ... & V+i\omega \\
... & ... & ... & ... & ... \\
V-i\omega & V-i\omega & V-i\omega & ... & 2V%
\end{array}%
\right) .    \notag \label{eq4-11}
\end{equation}
\begin{widetext}
For $\det M_{Vp}$ and $M_{Vp}^{-1},$ we have
\begin{eqnarray}
M_{Vp}^{-1} &=&\frac{1}{4}\left(
\begin{array}{ccccccc}
2V^{-1} & -(V^{-1}+i\omega ) & 0 & ... & 0 & 0 & 0 \\
-(V^{-1}-i\omega ) & 2V^{-1} & -(V^{-1}+i\omega ) & ... & 0 & 0 & 0 \\
0 & -(V^{-1}-i\omega ) & 2V^{-1} & ... & 0 & 0 & 0 \\
0 & 0 & -(V^{-1}-i\omega ) & ... & 0 & 0 & 0 \\
... & ... & ... & ... & ... & ... & ... \\
0 & 0 & 0 & ... & 2V & -(V^{-1}+i\omega ) & 0 \\
0 & 0 & 0 & ... & -(V^{-1}-i\omega ) & 2V^{-1} & -(V^{-1}+i\omega ) \\
0 & 0 & 0 & ... & 0 & -(V^{-1}-i\omega ) & 2V^{-1}%
\end{array}%
\right) ,  \label{eq4-12} \\
\det M_{Vp} &=&4^{n-1}.   \label{eq4-13}
\end{eqnarray}
\end{widetext}
From Eq. (\ref{eq4-3}), we can check that
\begin{eqnarray}
\det V &=&1,   \ \
V^{-1} =-\omega V\omega ,   \notag \label{eq4-14} \\
(V+i\omega )(V^{-1}-i\omega ) &=&0,  \ \
(V-i\omega )(V^{-1}+i\omega ) =0.    \notag \label{eq4-15}
\end{eqnarray}
With these facts we can directly verify Eq. (\ref{eq4-12}). $%
\{|\Phi _{j}\rangle \}_{j=1}^{n}$ are all pure, then the
symplectic eigenvalues of $V$ are all $1,$ together with
Eqs. (\ref{eq3-5},\ref{eq3-15}), we obtain Eq. (\ref{eq4-13}).

(1.a). For $n=2,$ $\zeta =0,$ Theorem \ref{Theorem-1} with Eqs. (\ref{eq4-3},\ref{eq4-10},\ref{eq4-12},\ref{eq4-13}) yields
\begin{equation}
\text{tr}\left( |\alpha _{1}\rangle \left\langle \alpha _{1}\right\vert
\alpha _{2}\rangle \left\langle \alpha _{2}\right\vert \right)
=|\left\langle \alpha _{1}\right\vert \alpha _{2}\rangle |^{2}=\exp
[-4|\alpha |^{2}].   \label{eq4-16}
\end{equation}%
Eq. (\ref{eq4-16}) can also be obtained by directly computing $|\left\langle \alpha
_{1}\right\vert \alpha _{2}\rangle |^{2}$ using Eqs. (\ref{eq4-5},\ref{eq4-9}).

Observe that
\begin{equation}
\{\exp [-4|\alpha |^{2}]:\alpha \in\mathbb{C}\}=(0,1].  \notag \label{eq4-17}
\end{equation}%
Meanwhile Eq. (\ref{eq3-4}) says $0\notin \mathcal{G}_{2,m}.$ Combining Eqs. (\ref{eq2-22},\ref{eq2-25}),
we thus get%
\begin{equation}
\mathcal{G}_{2,m}=(0,1], \ \text{for any} \ m\geq 1.  \label{eq4-18}
\end{equation}

(1.b). For $n\geq 3,$ direct computation (we give some details for Eq. (\ref{eq4-19}) in
Appendix \hyperlink{Appendix-C}{C}) by Theorem \ref{Theorem-1} with Eqs. (\ref{eq4-3},\ref{eq4-10},\ref{eq4-12},\ref{eq4-13}) yields
\begin{eqnarray}
&&\text{tr}\left( |\Phi _{1}\rangle \left\langle \Phi _{1}\right\vert \Phi
_{2}\rangle \left\langle \Phi _{2}\right\vert ...|\Phi _{n}\rangle
\left\langle \Phi _{n}\right\vert \right) \notag \\
&=&\exp \left\{ -2n|\alpha |^{2}\sin ^{2}\frac{\pi }{n}\left[ \cosh (2|\zeta
|)-i\cot \frac{\pi }{n}\right] \right\} .   \label{eq4-19}
\end{eqnarray}
We write tr$\left( |\Phi _{1}\rangle \left\langle \Phi _{1}\right\vert \Phi
_{2}\rangle \left\langle \Phi _{2}\right\vert ...|\Phi _{n}\rangle
\left\langle \Phi _{n}\right\vert \right) =re^{i\theta }$ in the polar form, and let $|\alpha|,$ $|\zeta|$ vary in $[0,\infty)$, hence
\begin{eqnarray}
r &=&\exp \left[ -2n|\alpha |^{2}\sin ^{2}\frac{\pi }{n}\cosh (2|\zeta |)%
\right] ,   \notag \label{eq4-20} \\
\theta  &=&2n|\alpha |^{2}\sin \frac{\pi }{n}\cos \frac{\pi }{n}\in \lbrack
0,\infty ),  \label{eq4-21} \\
r &=&\exp \left[ -\theta \tan \frac{\pi }{n}\cosh (2|\zeta |)\right] .  \label{eq4-22}
\end{eqnarray}%
Note that $r$ and $\theta $ are all independent of $\phi .$

Since
\begin{equation}
\{\cosh (2|\zeta |):\zeta \in\mathbb{C}\}=[1,\infty ).  \label{eq4-23}
\end{equation}%
Eqs. (\ref{eq4-21},\ref{eq4-22},\ref{eq4-23}) then determine a region $\mathcal{E}_{n}^{\prime }\subseteq
\mathcal{G}_{n,1}$ in the complex plane as
\begin{equation}
\mathcal{E}_{n}^{\prime }=\{e^{i\theta }e^{-t\theta \tan \frac{\pi }{n}%
}:\theta \in \lbrack 0,+\infty ),t\in \lbrack 1,\infty )\}.   \label{eq4-24}
\end{equation}%
When $\zeta =0,$ then $\cosh (2|\zeta |)=1,$ $\mathcal{E}_{n}^{\prime }$
degenerates into
\begin{equation}
\mathcal{E}_{n}^{^{\prime \prime }}=\{e^{i\theta }e^{-\theta \tan \frac{\pi
}{n}}:\theta \in \lbrack 0,\infty )\}.   \label{eq4-25}
\end{equation}%
$\mathcal{E}_{n}^{^{\prime \prime }}$ represents a logarithmic spiral in the
complex plane. clearly, $e^{-\theta \tan \frac{\pi }{n}}$ strictly decreases
with $\theta $.

Notice that in Eq. (\ref{eq4-24}),
\begin{eqnarray}
\{e^{i\theta }e^{-t\theta \tan \frac{\pi }{n}} &:&\theta =0,t\in \lbrack
1,+\infty )\}=\{1\}; \notag \\
\{e^{i\theta }e^{-t\theta \tan \frac{\pi }{n}} &:&\theta =2\pi ,t\in \lbrack
1,+\infty )\}=(0,e^{-2\pi \tan \frac{\pi }{n}}]. \notag
\end{eqnarray}
That is,
\begin{equation}
(e^{-2\pi \tan \frac{\pi }{n}},1)\cap \mathcal{E}_{n}^{\prime }=\varnothing .   \notag
\end{equation}

As a special case (pure states) of Eq. (\ref{eq2-20}), we have
\begin{eqnarray}
&&[\text{tr}\left( |\Phi _{1}\rangle \left\langle \Phi _{1}\right\vert \Phi
_{2}\rangle \langle \Phi _{2}|...|\Phi _{n}\rangle \left\langle \Phi
_{n}\right\vert \right) ]^{\ast } \notag \\
&=&\text{tr}\left( |\Phi _{n}\rangle \left\langle \Phi _{n}\right\vert
...|\Phi _{2}\rangle \langle \Phi _{2}|\Phi _{1}\rangle \left\langle \Phi
_{1}\right\vert \right) =re^{-i\theta },  \notag
\end{eqnarray}%
this leads to the sets $\mathcal{E}_{n}^{\prime \ast }$ and $\mathcal{E}%
_{n}^{^{\prime \prime }\ast }$ by Eqs. (\ref{eq2-24},\ref{eq2-25}).

(1.c). For $n\geq 3,$ we consider the ordered set of one-mode Glauber
coherent states $\{|\alpha _{j}\rangle \}_{j=1}^{n}$ with $\alpha
_{1}=|\alpha |,$ $\alpha _{2}=\alpha _{3}=...=\alpha _{n}=-|\alpha |,$ then Eq. (\ref{eq4-5}) yields
\begin{eqnarray}
&&\text{tr}\left( |\alpha _{1}\rangle \left\langle \alpha _{1}\right\vert
\alpha _{2}\rangle \left\langle \alpha _{2}\right\vert ...|\alpha
_{n}\rangle \left\langle \alpha _{n}\right\vert \right)  \notag  \\
&=&|\left\langle \alpha _{1}\right\vert \alpha _{2}\rangle |^{2}=\exp
[-4|\alpha |^{2}].   \notag
\end{eqnarray}%
Similar to (1.a), we have
\begin{equation}
(0,1]\subseteq \mathcal{G}_{n,1}.    \notag \label{eq4-26}
\end{equation}

For $n\geq 3,$ in (1.b) and (1.c), we have obtained the subsets of $\mathcal{%
G}_{n,1}:\mathcal{E}_{n}^{\prime }$, $\mathcal{E}_{n}^{\prime \ast }$ and $%
(0,1]$. Now we define the set
\begin{equation}
\mathcal{E}_{n}=\mathcal{E}_{n}^{\prime }\cup \mathcal{E}_{n}^{\prime \ast
}\cup (0,1].    \notag \label{eq4-27}
\end{equation}
Evidently,
\begin{equation}
\mathcal{E}_{n}=\{te^{i\theta }e^{-\theta g(\theta )\tan \frac{\pi }{n}%
}:\theta \in \lbrack -\pi ,\pi ],t\in (0,1]\},    \label{eq4-28}
\end{equation}%
where $g(\theta )$ is the sign function of $\theta ,$ namely, $g(\theta )=1$
if $\theta >0,$ $g(\theta )=-1$ if $\theta <0,$ and $g(\theta )=0$ if $%
\theta =0.$ Note that $0\notin \mathcal{E}_{n}.$

We denote the boundary of $\mathcal{E}_{n}$ by $\partial \mathcal{E}_{n},$
that is,
\begin{equation}
\partial \mathcal{E}_{n}=\{e^{i\theta }e^{-\theta g(\theta )\tan \frac{\pi }{%
n}}:\theta \in \lbrack -\pi ,\pi ]\}.   \label{eq4-29}
\end{equation}%
$\partial \mathcal{E}_{n}$ can be viewed as the subset of $\mathcal{E}_{n}$
by letting $t=1$ in $\mathcal{E}_{n}.$ Conversely, $\mathcal{E}_{n}$ is the
region enclosed by $\partial \mathcal{E}_{n}$ ($\mathcal{E}_{n}$ contains $%
\partial \mathcal{E}_{n}),$ except $0\notin \mathcal{E}_{n}.$ Clearly,
\begin{equation}
\mathcal{E}_{n}=\mathcal{E}_{n}^{\ast }, \ \
\partial \mathcal{E}_{n}=(\partial \mathcal{E}_{n})^{\ast }.    \notag \label{eq4-30}
\end{equation}

Comparing Eq. (\ref{eq4-25}) with Eq. (\ref{eq2-23}), we can prove that (we give a proof for Eq. (\ref{eq4-31}) in
Appendix \hyperlink{Appendix-E}{E})
\begin{equation}
e^{-\theta \tan \frac{\pi }{n}}\leq \cos ^{n}\frac{\pi }{n}\sec ^{n}\frac{%
\pi -\theta }{n}, \ \theta \in \lbrack 0,\pi ],   \label{eq4-31}
\end{equation}%
and equality holds if and only if $\theta =0.$ This shows $\mathcal{E}%
_{n}\varsubsetneqq \mathcal{B}_{n}.$

\begin{figure*}[htbp]   \hypertarget{Figure-1}{}
  \begin{minipage}[t]{0.65\textwidth}
     \begin{minipage}[t]{1.0\textwidth}
     \includegraphics[width=0.45\textwidth]{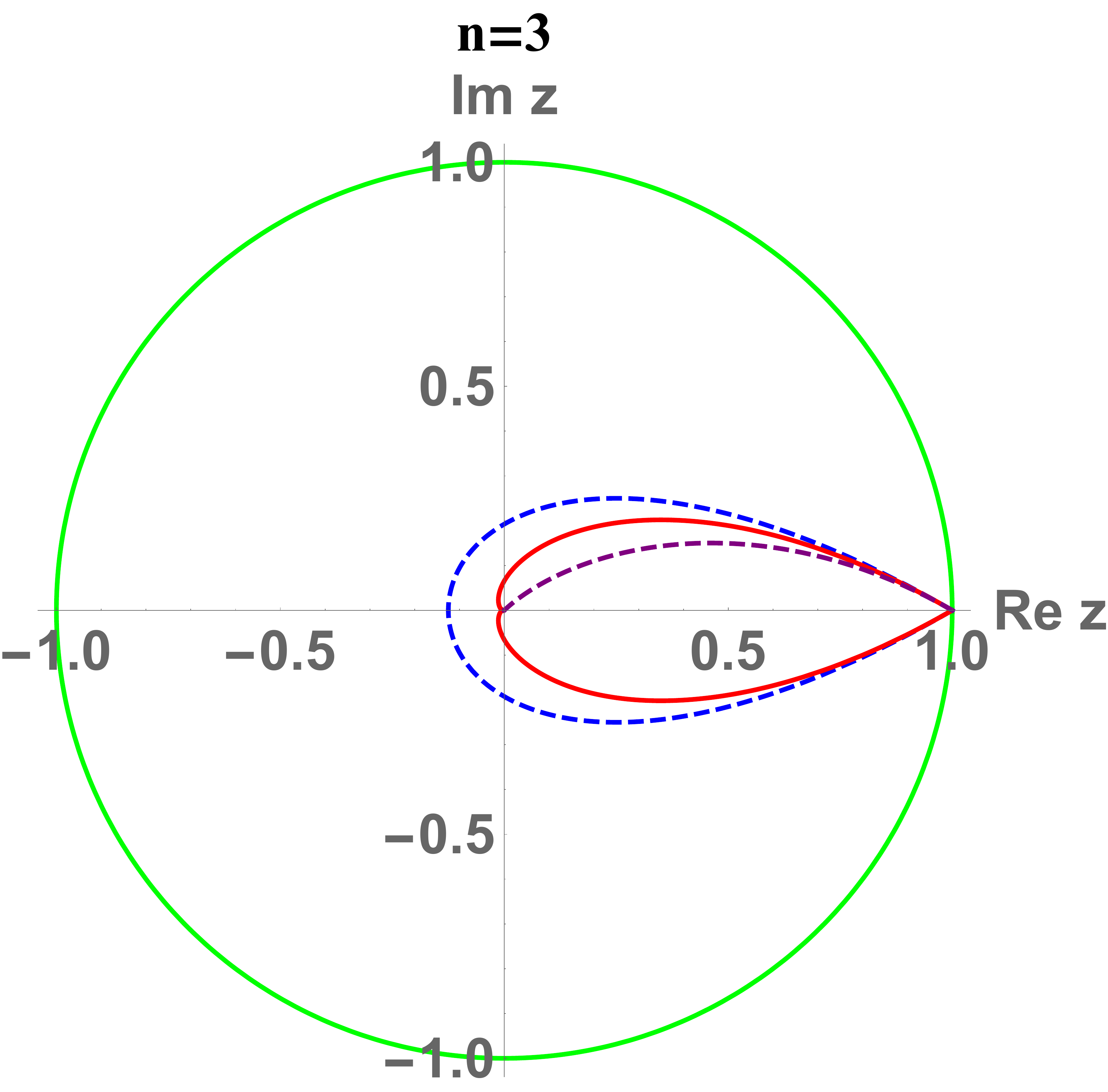} \ \ \
     \includegraphics[width=0.45\textwidth]{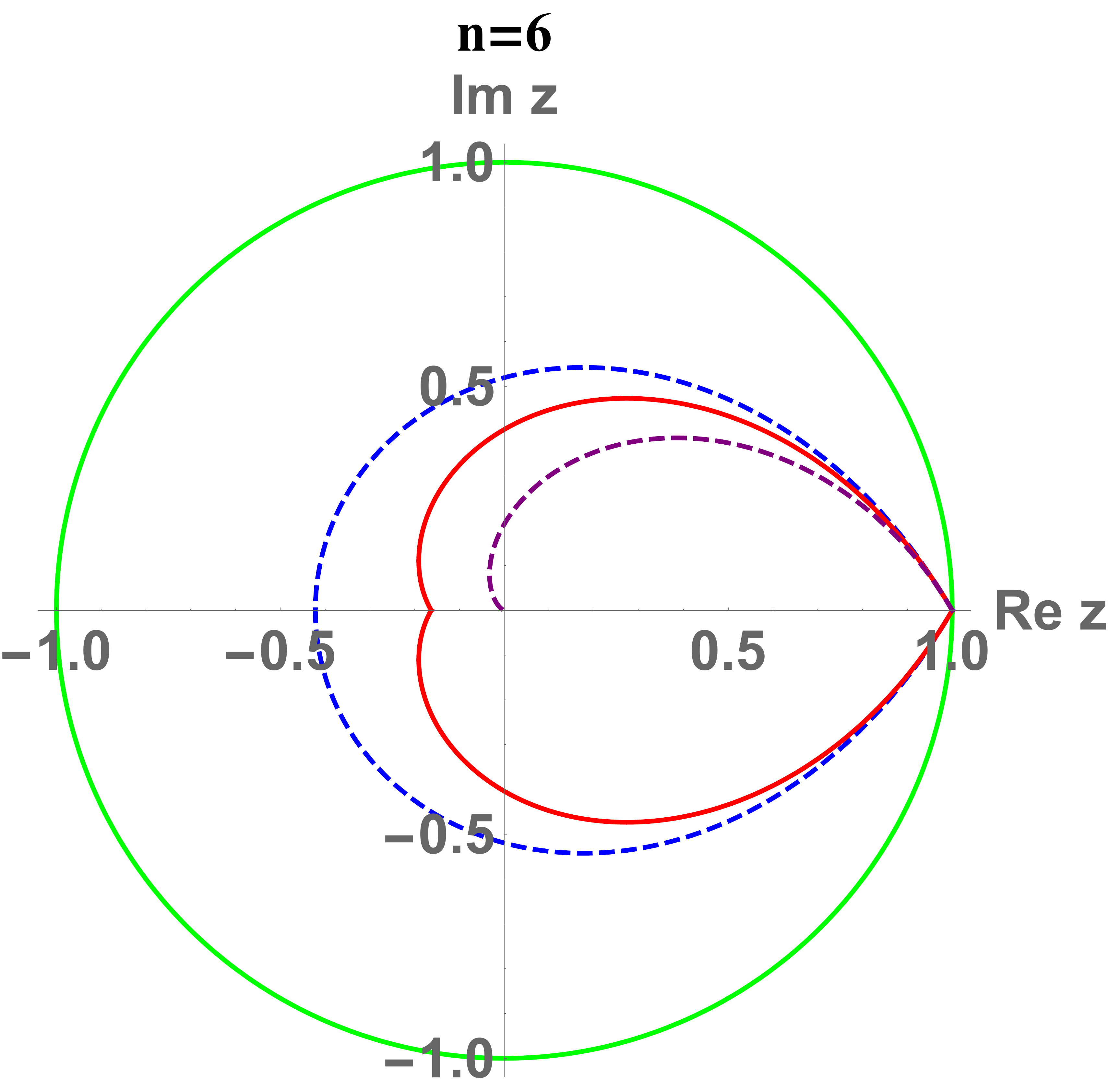}
     \end{minipage}
     \vspace{0.2cm} \\
     \begin{minipage}[t]{1.0\textwidth}
     \includegraphics[width=0.45\textwidth]{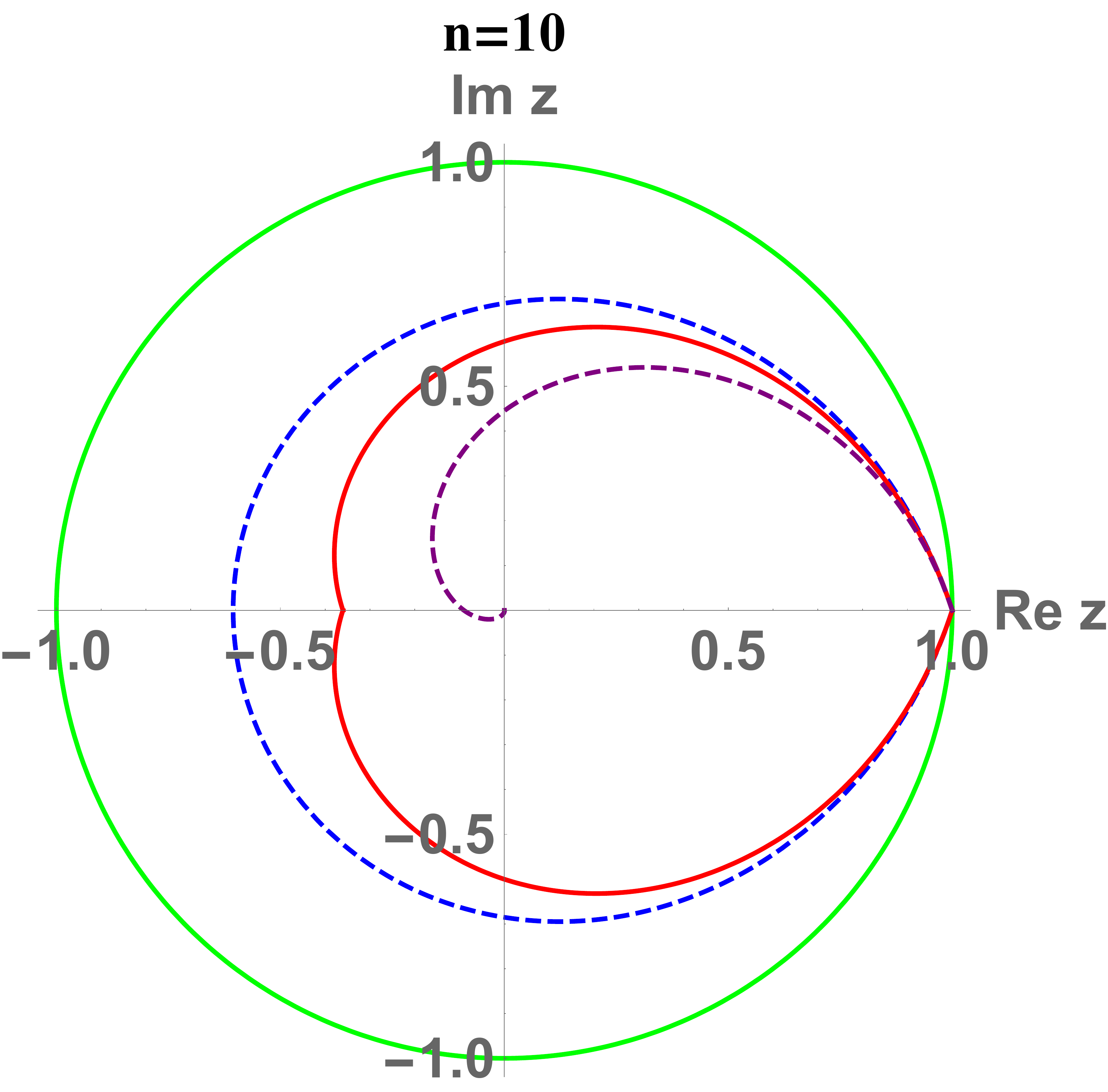} \ \ \
     \includegraphics[width=0.45\textwidth]{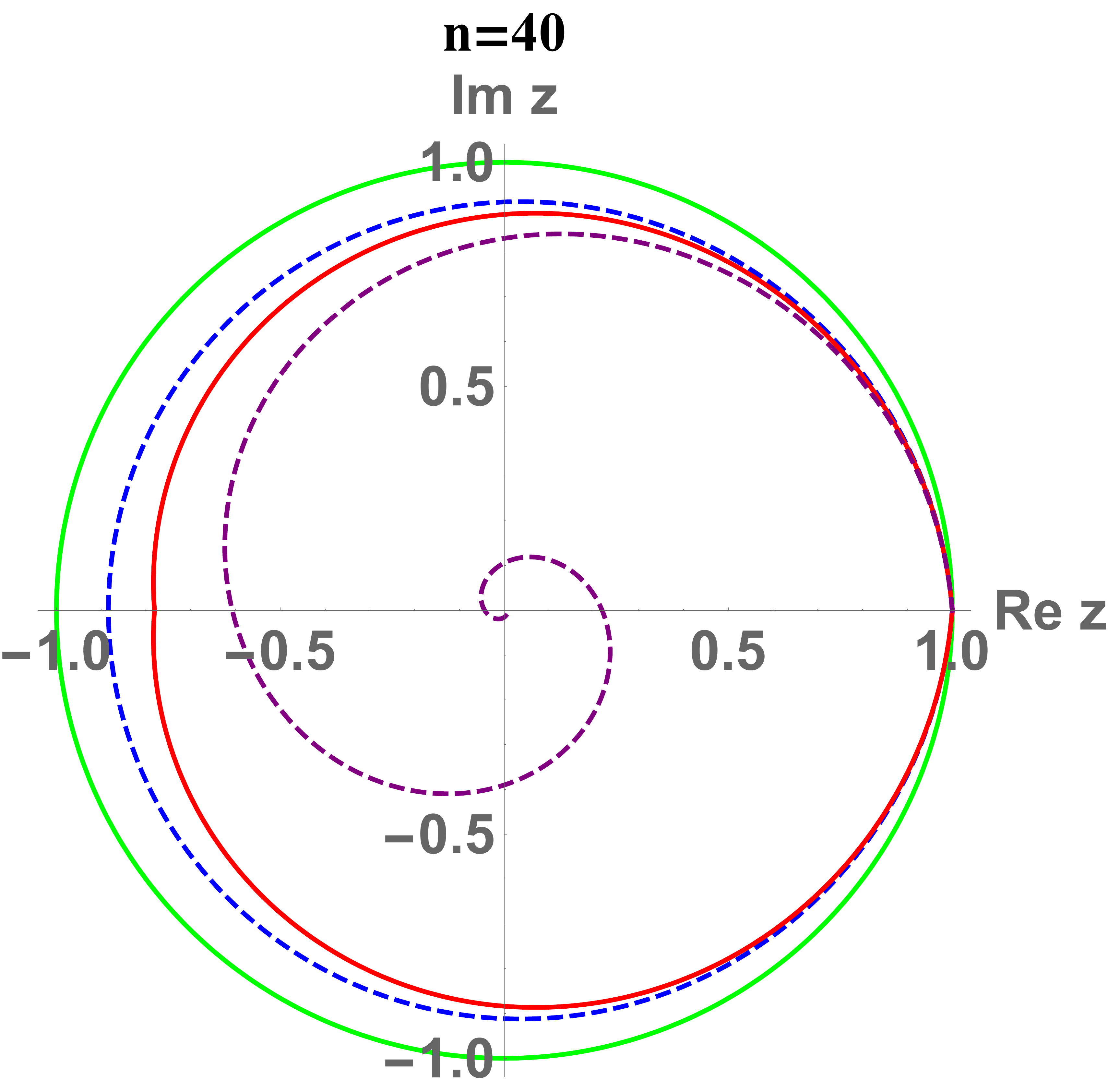}
     \end{minipage}
  \end{minipage}
  \hspace{-1.2cm}
  \begin{minipage}[t]{0.32\textwidth}
  \vspace{-3.2cm}
  \includegraphics[width=1.2\textwidth]{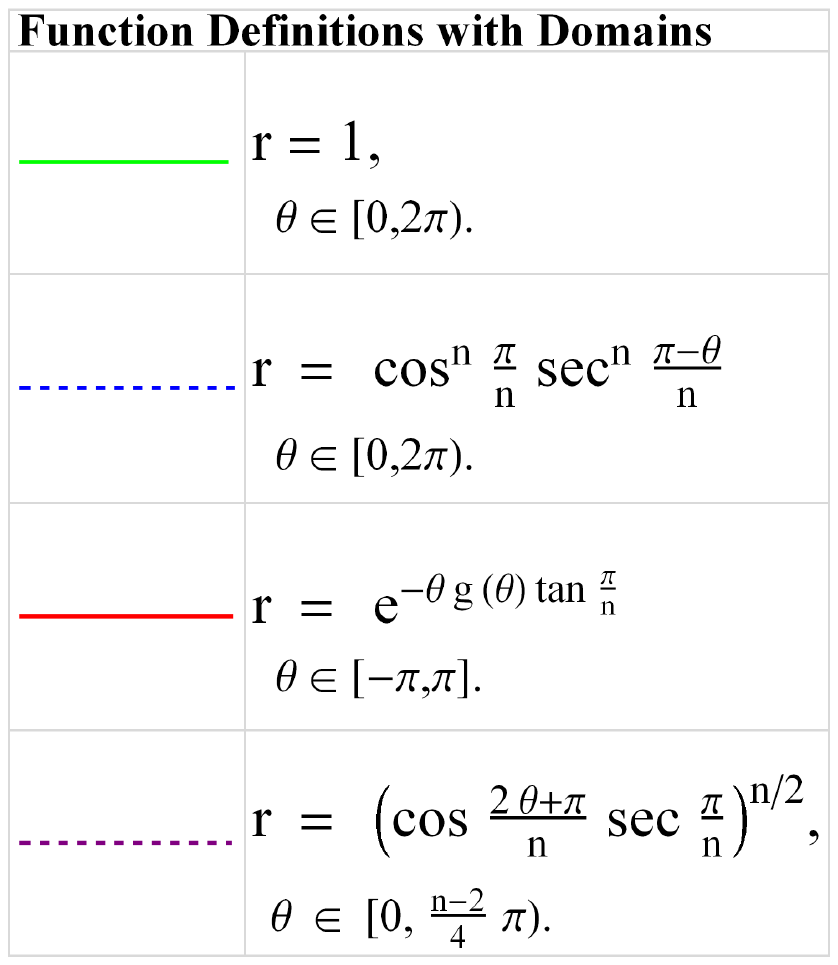}
  \end{minipage}
  \caption{Comparison of $\mathcal{F}_{n}$ (purple dashed curve) in Eq. (\ref{eq4-41}), $\partial \mathcal{E}_{n}$ (red solid curve) in Eq. (\ref{eq4-29}), and $\partial\mathcal{B}_{n}$ (blue dashed curve) in Eq. (\ref{eq2-23-1}), for $n=3,6,10,40.$ All curves are confined within the unit circle (green solid curve), and $(\mathcal{F}_{n}\cup \mathcal{F}_{n}^{\ast })\subsetneqq \mathcal{E}_{n}\subsetneqq \mathcal{B}_{n}$ holds in Eq. (\ref{eq4-43}). In the complex plane, we write any complex number $z$ in the polar form $z=re^{i\theta}$. }
\end{figure*}

\textbf{Example 2.} Consider the ordered set of one-mode squeezed states $\{|\zeta
_{j}\rangle \}_{j=1}^{n},$
\begin{equation}
\zeta _{j}=|\zeta |\exp \left( i\frac{2\pi }{n}j\right) .  \label{eq4-32}
\end{equation}%
The mean of $|\zeta _{j}\rangle $ is $\overline{X}^{(j)}=(0,0)^{T},$ the
covariance matrix $V^{(j)}$ of $|\zeta _{j}\rangle $ is
\begin{equation}
\begin{cases}
V_{11}^{(j)}=\cosh (2|\zeta |)+\cos \left( \frac{2\pi }{n}j\right) \sinh
(2|\zeta |) \\
V_{12}^{(j)}=V_{21}^{(j)}=\sin \left( \frac{2\pi }{n}j\right) \sinh (2|\zeta
|) \\
V_{22}^{(j)}=\cosh (2|\zeta |)-\cos \left( \frac{2\pi }{n}j\right) \sinh
(2|\zeta |).%
\end{cases} \label{eq4-33}
\end{equation}

Eq. (\ref{eq3-3}) yields $\Lambda =0.$ By Theorem \ref{Theorem-1},
\begin{equation}
\text{tr}\left( |\zeta_{1}\rangle \left\langle \zeta
_{1}\right\vert \zeta_{2}\rangle \langle \zeta_{2}|...|\zeta_{n}\rangle
\left\langle \zeta_{n}\right\vert \right) =\frac{2^{n-1}}{\sqrt{\det M}}, \label{eq4-34}
\end{equation}%
$M$ is defined by Eqs. (\ref{eq3-2},\ref{eq4-33}). For one-mode squeezed states $\{|\zeta
_{j}\rangle \}_{j=1}^{n}$ defined in Eq. (\ref{eq4-32}), tr$\left( |\zeta_{1}\rangle \left\langle \zeta_{1}\right\vert ...|\zeta_{n}\rangle \left\langle \zeta
_{n}\right\vert \right)$ can be
easily computed by Eq. (\ref{eq4-7}), that is
\begin{eqnarray}
\text{tr}\left( |\zeta_{1}\rangle \left\langle \zeta
_{1}\right\vert \zeta_{2}\rangle \langle \zeta_{2}|...|\zeta_{n}\rangle
\left\langle \zeta_{n}\right\vert \right)
=\frac{1}{\left( \sqrt{\cosh ^{2}|\zeta |-e^{i\frac{2\pi }{n}}\sinh
^{2}|\zeta |}\right) ^{n}}.  \notag \\  \label{eq4-35}
\end{eqnarray}
Comparing Eqs. (\ref{eq4-34},\ref{eq4-35}), one gets
\begin{equation}
\det M=4^{n-1}\left( \cosh ^{2}|\zeta |-e^{i\frac{2\pi }{n}}\sinh ^{2}|\zeta
|\right) ^{n}.    \notag \label{eq4-36}
\end{equation}

For $n\geq 2,$ we define the sets
\begin{eqnarray}
\mathcal{F}_{n} &=&\{\text{tr}\left( |\zeta_{1}\rangle \left\langle \zeta
_{1}\right\vert \zeta_{2}\rangle \langle \zeta_{2}|...|\zeta_{n}\rangle
\left\langle \zeta_{n}\right\vert \right) :\zeta \in\mathbb{C}\},   \notag \label{eq4-37}  \\
\mathcal{F}_{n}^{\ast } &=&\{\text{tr}\left( |\zeta_{n}\rangle \left\langle
\zeta_{n}\right\vert \zeta_{n-1}\rangle \langle \zeta_{n-1}|...|\zeta
_{1}\rangle \left\langle \zeta_{1}\right\vert \right) :\zeta \in\mathbb{C}\}.   \notag  \label{eq4-38}  \ \ \ \
\end{eqnarray}

(2.a). When $n=2,$ Eq. (\ref{eq4-35}) yields
\begin{equation}
\text{tr}\left( |\zeta_{1}\rangle \left\langle \zeta_{1}\right\vert \zeta
_{2}\rangle \langle \zeta _{2}|\right) =\frac{1}{\cosh (2|\zeta |)}.    \notag \label{eq4-39}
\end{equation}%
With Eq. (\ref{eq4-23}), one has
\begin{equation}
\mathcal{F}_{2}=\mathcal{F}_{2}^{\ast }=\left\{ \frac{1}{\cosh (2|\zeta |)}%
:\zeta \in\mathbb{C}\right\} =(0,1].    \notag \label{eq4-40}
\end{equation}%
With Eq. (\ref{eq2-26}), we see that tr$\left( |\Psi _{1}\rangle \left\langle \Psi _{1}\right\vert \Psi
_{2}\rangle \langle \Psi _{2}|\right) $ with $\zeta \in\mathbb{C}$ also realizes Eq. (\ref{eq4-18}).

(2.b) When $n\geq 3,$ in Appendix \hyperlink{Appendix-D}{D} we will prove
\begin{equation}
\mathcal{F}_{n}=\left\{ e^{i\theta }\left( \cos \frac{2\theta +\pi }{n}\sec
\frac{\pi }{n}\right) ^{\frac{n}{2}}:\theta \in \left[ 0,\frac{n-2}{4}\pi
\right) \right\} .   \label{eq4-41}
\end{equation}%
Since $\theta \in \left[ 0,\frac{n-2}{4}\pi \right) ,$ then $\frac{2\theta
+\pi }{n}\in \left[ \frac{\pi }{n},\frac{\pi }{2}\right) $ and $\left( \cos
\frac{2\theta +\pi }{n}\sec \frac{\pi }{n}\right) ^{\frac{n}{2}}$ strictly
decreases with $\theta $.

Comparing Eq. (\ref{eq4-41}) with Eq. (\ref{eq4-29}), we can prove that (we give a proof for Eq. (\ref{eq4-42}) in
Appendix \hyperlink{Appendix-E}{E})
\begin{equation}
\left( \cos \frac{2\theta +\pi }{n}\sec \frac{\pi }{n}\right) ^{%
\frac{n}{2}}\leq e^{-\theta \tan \frac{\pi }{n}}, \theta \in
\left[0,\min \left\{\frac{n-2}{4}\pi ,\pi \right\}\right),   \label{eq4-42}
\end{equation}%
and equality holds if and only if $\theta =0.$ As a result,
\begin{equation}
(\mathcal{F}_{n}\cup \mathcal{F}_{n}^{\ast })\subsetneqq \mathcal{E}%
_{n}\subsetneqq \mathcal{B}_{n}.     \label{eq4-43}
\end{equation}

We depict $\mathcal{F}_{n},$ $\partial \mathcal{E}_{n}$ and $\partial\mathcal{B}_{n}$
for $n=3,$ $6,$ $10,$ $40$ in Figure \hyperlink{Figure-1}{1}.

In this section, we have shown $\mathcal{G}_{2,m}=(0,1]$ and have successfully constructed a subset $\mathcal{E}_{n}$ for $\mathcal{G}_{n,m}$ for any $n \geq 3$. While this represents a key step forward, a complete characterization of $\mathcal{G}_{n,m}$ for arbitrary $n \geq 3$ and $m \geq 1$ remains a significant open problem in theoretical research. To this problem we have the following conjecture.
\begin{Conjecture} \label{Conjecture-1}
\begin{equation}
\mathcal{E}_{n}=\mathcal{G}_{n,m} \ \text {for any} \ n\geq 3 \ \text {and} \ m\geq 1. \notag
\end{equation}
\end{Conjecture}

\hypertarget{section V}{}
\section{Summary}
The Bargmann invariant is a fundamental quantity in quantum mechanics with far-reaching theoretical and experimental implications. In this work, we have derived a general expression for the Bargmann invariant of arbitrary
$m$-mode Gaussian states, explicitly formulated in terms of their means and covariance matrices. This result provides a rigorous foundation for further investigations into the related properties of Gaussian states, opening new avenues for both theoretical explorations and experimental validations.

As a key finding, we have identified a restricted set of permissible values for the Bargmann invariants of Gaussian states, denoted by $\mathcal{E}_{n}$ for any $n\geq 3$ in Eq. (\ref{eq4-28}), which can be realized by only one-mode pure Gaussian states. A central theoretical challenge that emerges from our work is the complete characterization of permissible Bargmann invariant values for general $m$-mode Gaussian states, i.e., the set $\mathcal{G}_{n,m}$ in Eq. (\ref{eq2-24}). We conjecture that $\mathcal{E}_{n}=\mathcal{G}_{n,m}$ for any $n\geq 3$ and $m\geq 1$. Resolving this challenge would not only deepen our understanding of Gaussian states, but also have potential applications in quantum information processing.

\begin{widetext}
\section*{ACKNOWLEDGMENTS}

This work is supported by the National Natural Science Foundation of China (Grant No. 12471443).

\hypertarget{Appendix-A}{}

\section*{Appendix A: proof of Theorem \ref{Theorem-1}}
\setcounter{equation}{0} \renewcommand%
\theequation{A\arabic{equation}}

We first state a Lemma on Gaussian integral.
\begin{Lemma} \label{Lemma-1}
(Gaussian integral formula, see for example \cite{Folland-1989-book}) It holds that
\begin{eqnarray}
\int_{-\infty }^{\infty }dx_{1}\int_{-\infty }^{\infty
}dx_{2}...\int_{-\infty }^{\infty }dx_{n}\exp \left[-\frac{1}{2}X^{T}QX+\Lambda
^{T}X\right]=\frac{\sqrt{(2\pi )^{n}}}{\sqrt{\det Q}}\exp \left[\frac{1}{2}\Lambda
^{T}Q^{-1}\Lambda \right],   \notag \label{eqA-1}
\end{eqnarray}
where $X=(x_{1},x_{2},...,x_{n})^{T}$ is $n$-dimensional real column vector, $Q=Q^{T}$ is a symmetric complex matrix, $\Lambda $ is $n$-dimensional complex vector, Re$Q$ is positive definite, $\sqrt{\det Q}$
takes the principal branch of square root function $\sqrt{z}$ such that $\sqrt{z}>0$
when $z>0.$ Re$Q=\frac{1}{2}(Q+Q^{\ast })$ is the real part of $Q.$
\end{Lemma}

For $\rho _{j}(\overline{X}^{(j)},V^{(j)}),$ we introduce $\xi ^{(j)}=(\xi
_{1}^{(j)},\xi _{2}^{(j)},...,\xi _{2m}^{(j)})^{T}\in\mathbb{R}
^{2m}$ similarly to $\xi $ defined in Eqs. (\ref{eq2-10}), and let $\lambda
_{k}^{(j)}=(\xi _{2k-1}^{(j)},\xi _{2k}^{(j)})^{T}$ for $1\leq k\leq m.$
From Eq. (\ref{eq2-15}), we have
\begin{equation}
\rho _{j}=\int \frac{d^{2m}\xi ^{(j)}}{\pi ^{m}}\chi (\rho ,\xi
^{(j)})D(-\xi ^{(j)}), \ \ 1\leq j\leq n.  \label{eqA-2}
\end{equation}
Eq. (\ref{eq2-9}) implies
\begin{eqnarray}
D(\xi ^{(j)})=D(\lambda _{1}^{(j)})D(\lambda_{2}^{(j)})...D(\lambda _{m}^{(j)}).  \label{eqA-3}
\end{eqnarray}
Then Eqs. (\ref{eqA-2}) lead to
\begin{eqnarray}
\text{tr}(\rho _{1}\rho _{2}...\rho _{n})
&=&\text{tr}\left[ \int \frac{d^{2m}\xi ^{(1)}}{\pi ^{m}}\chi (\rho _{1},\xi
^{(1)})D(-\xi ^{(1)})\int \frac{d^{2m}\xi ^{(2)}}{\pi ^{m}}\chi (\rho
_{2},\xi ^{(2)})D(-\xi ^{(2)})...\int \frac{d^{2m}\xi ^{(n)}}{\pi ^{m}}\chi
(\rho _{n},\xi ^{(n)})D(-\xi ^{(n)})\right]   \notag \\
&=&\int \frac{d^{2m}\xi ^{(1)}}{\pi ^{m}}\int \frac{d^{2m}\xi ^{(2)}}{\pi
^{m}}...\int \frac{d^{2m}\xi ^{(n)}}{\pi ^{m}}\chi (\rho _{1},\xi
^{(1)})\chi (\rho _{2},\xi ^{(2)})...\chi (\rho _{n},\xi ^{(n)})\text{tr}%
[D(-\xi ^{(1)})D(-\xi ^{(2)})...D(-\xi ^{(n)})].    \label{eqA-4}
\end{eqnarray}
In Eq. (\ref{eqA-4}), using Eq. (\ref{eqA-3}), we find
\begin{eqnarray}
&&\text{tr}[D(-\xi ^{(1)})D(-\xi ^{(2)})...D(-\xi ^{(n)})]     \notag   \\
&=&\text{tr}\{[D(-\lambda _{1}^{(1)})D(-\lambda _{2}^{(1)})...D(-\lambda
_{m}^{(1)})][D(-\lambda _{1}^{(2)})D(-\lambda _{2}^{(2)})...D(-\lambda
_{m}^{(2)})]...[D(-\lambda _{1}^{(n)})D(-\lambda _{2}^{(n)})...D(-\lambda
_{m}^{(n)})]\}     \notag   \\
&=&\text{tr}\{[D(-\lambda _{1}^{(1)})D(-\lambda _{1}^{(2)})...D(-\lambda
_{1}^{(n)})][D(-\lambda _{2}^{(1)})D(-\lambda _{2}^{(2)})...D(-\lambda
_{2}^{(n)})]...[D(-\lambda _{m}^{(1)})D(-\lambda _{m}^{(2)})...D(-\lambda
_{m}^{(n)})]\}       \notag \\
&=&\{\text{tr}[D(-\lambda _{1}^{(1)})D(-\lambda _{1}^{(2)})...D(-\lambda
_{1}^{(n)})]\}\{\text{tr}[D(-\lambda _{2}^{(1)})D(-\lambda
_{2}^{(2)})...D(-\lambda _{2}^{(n)})]\}...\{\text{tr}[D(-\lambda
_{m}^{(1)})D(-\lambda _{m}^{(2)})...D(-\lambda _{m}^{(n)})]\}.     \label{eqA-5}
\end{eqnarray}

In Eq. (\ref{eqA-5}), we have \cite{Serafini-2023-book}
\begin{eqnarray}
D(\lambda _{k}^{(1)})D(\lambda _{k}^{(2)})=D\left(
\lambda _{k}^{(1)}+\lambda _{k}^{(2)}\right) \exp \left( -i\lambda _{k}^{(1)T}\omega \lambda _{k}^{(2)}\right),   \notag  \label{eqA-6}
\end{eqnarray}
this yields
\begin{eqnarray}
D(\lambda _{k}^{(1)})D(\lambda _{k}^{(2)})...D(\lambda _{k}^{(n)})=D\left(
\sum_{j=1}^{n}\lambda _{k}^{(j)}\right) \exp \left( -i\sum_{1\leq
j<j^{\prime }\leq n}\lambda _{k}^{(j)T}\omega \lambda _{k}^{(j^{\prime
})}\right) .    \label{eqA-7}
\end{eqnarray}

For $D(\lambda _{k}^{(j)}),$ it is the fact that \cite{Glauber-1969-PR}
\begin{equation}
\text{tr}D(\lambda _{k}^{(j)})=\pi \delta (\lambda _{k}^{(j)}),    \label{eqA-8}
\end{equation}
where $\delta (\lambda _{k}^{(j)})=\delta (\xi _{2k-1}^{(j)})\delta (\xi
_{2k}^{(j)})$ is the delta function in $\mathbb{R}^{2}.$

Taking Eqs. (\ref{eqA-7},\ref{eqA-8}) into Eq. (\ref{eqA-5}), one gets
\begin{eqnarray}
\text{tr}[D(-\lambda _{k}^{(1)})D(-\lambda _{k}^{(2)})...D(-\lambda
_{k}^{(n)})] &=&\text{tr}[D(\lambda _{k}^{(1)})D(\lambda _{k}^{(2)})...D(\lambda
_{k}^{(n)})]=\pi \delta \left( \sum_{j=1}^{n}\lambda _{k}^{(j)}\right)
\exp \left( -i\sum_{1\leq j<j^{\prime }\leq n}\lambda _{k}^{(j)T}\omega
\lambda _{k}^{(j^{\prime })}\right) ,      \notag \label{eqA-9}  \\
\text{tr}[D(-\xi ^{(1)})D(-\xi ^{(2)})...D(-\xi ^{(n)})] &=&\pi
^{m}\prod_{k=1}^{m}\delta \left( \sum_{j=1}^{n}\lambda _{k}^{(j)}\right)
\exp \left( -i\sum_{1\leq j<j^{\prime }\leq n}\lambda _{k}^{(j)T}\omega
\lambda _{k}^{(j^{\prime })}\right) .    \label{eqA-10}
\end{eqnarray}

Delta function has the property
\begin{equation}
\int_{-\infty }^{\infty }f(x)\delta \left( x-x_{0}\right) dx=f(x_{0})     \label{eqA-11}
\end{equation}
for continuous function $f(x)$ and $x_{0}\in
\mathbb{R}.$

Substituting Eqs. (\ref{eqA-10},\ref{eqA-11}) into Eq. (\ref{eqA-4}), and integrating with respect to $\xi ^{(n)}$ in Eq. (\ref{eqA-4}), we obtain
\begin{eqnarray}
\text{tr}(\rho _{1}\rho _{2}...\rho _{n})
=\int \frac{d^{2m}\xi ^{(1)}}{\pi ^{m}}\int \frac{d^{2m}\xi ^{(2)}}{\pi
^{m}}...\int \frac{d^{2m}\xi ^{(n-1)}}{\pi ^{m}}\chi (\rho _{1},\xi
^{(1)})\chi (\rho _{2},\xi ^{(2)})...\chi (\rho _{n},\xi
^{(n)})\prod_{k=1}^{m}\exp \left( -i\sum_{1\leq j<j^{\prime }\leq n}\lambda
_{k}^{(j)T}\omega \lambda _{k}^{(j^{\prime })}\right) ,    \label{eqA-12}
\end{eqnarray}
where
\begin{eqnarray}
\xi ^{(n)}=-\sum_{j=1}^{n-1}\xi ^{(j)},   \ \ \ \
\lambda _{k}^{(n)}=-\sum_{j=1}^{n-1}\lambda _{k}^{(j)},1\leq k\leq m.    \label{eqA-13}
\end{eqnarray}

Employing Eqs. (\ref{eq2-14},\ref{eqA-13}), in Eq. (\ref{eqA-12}), $\chi (\rho _{n},\xi ^{(n)})$ reads
\begin{eqnarray}
\chi (\rho _{n},\xi ^{(n)}) &=&\exp \left[ -\frac{1}{2}\xi ^{(n)T}(\Omega
V^{(n)}\Omega ^{T})\xi ^{(n)}-i(\Omega \overline{X}^{(n)})^{T}\xi ^{(n)}%
\right]   \notag \\
&=&\exp \left[ -\frac{1}{2}\left( -\sum_{j=1}^{n-1}\xi ^{(j)}\right)
^{T}(\Omega V^{(n)}\Omega ^{T})\left( -\sum_{j=1}^{n-1}\xi ^{(j)}\right)
-i(\Omega \overline{X}^{(n)})^{T}\left( -\sum_{j=1}^{n-1}\xi ^{(j)}\right) %
\right]      \notag   \\
&=&\exp \left[ -\frac{1}{2}\sum_{j,j^{\prime }=1}^{n-1}\xi ^{(j)T}(\Omega
V^{(n)}\Omega ^{T})\xi ^{(j^{\prime })}+i\sum_{j=1}^{n-1}(\Omega \overline{X}%
^{(n)})^{T}\xi ^{(j)}\right].    \label{eqA-14}
\end{eqnarray}

From the definition of $\omega $ in Eq. (\ref{eq2-7}), we see that $\omega ^{T}=-\omega
.$ Consequently, in Eq. (\ref{eqA-12}),
\begin{eqnarray}
\lambda _{k}^{(j)T}\omega \lambda _{k}^{(j^{\prime })}=-\lambda
_{k}^{(j^{\prime })T}\omega \lambda _{k}^{(j)}, \ \ \ \
\lambda _{k}^{(j)T}\omega\lambda _{k}^{(j)}=0,       \label{eqA-15}
\end{eqnarray}
for any $1\leq j\leq n$ and any $1\leq j^{\prime }\leq n.$ Thus in Eq. (\ref{eqA-12}), Eqs. (\ref{eqA-13},\ref{eqA-15}) yield
\begin{eqnarray}
\sum_{1\leq j<j^{\prime }\leq n}\lambda _{k}^{(j)T}\omega \lambda
_{k}^{(j^{\prime })} &=&\sum_{1\leq j<j^{\prime }\leq n-1}\lambda
_{k}^{(j)T}\omega \lambda _{k}^{(j^{\prime })}+\sum_{1\leq j\leq n-1}\lambda
_{k}^{(j)T}\omega \lambda _{k}^{(n)}    \notag   \\
&=&\sum_{1\leq j<j^{\prime }\leq n-1}\lambda _{k}^{(j)T}\omega \lambda
_{k}^{(j^{\prime })}-\left( \sum_{j=1}^{n-1}\lambda _{k}^{(j)}\right)
^{T}\omega \left( \sum_{j=1}^{n-1}\lambda _{k}^{(j)}\right)  \notag   \\
&=&\sum_{1\leq j<j^{\prime }\leq n-1}\lambda _{k}^{(j)T}\omega \lambda
_{k}^{(j^{\prime })}  \notag   \\
&=&(\lambda _{k}^{(1)T},\lambda _{k}^{(2)T},...,\lambda _{k}^{(n-1)T})\left(
\begin{array}{cccccc}
0 & \omega & \omega & ... & \omega & \omega \\
0 & 0 & \omega & ... & \omega & \omega \\
0 & 0 & 0 & ... & \omega & \omega \\
... & ... & ... & ... & ... & ... \\
0 & 0 & 0 & ... & 0 & \omega \\
0 & 0 & 0 & ... & 0 & 0%
\end{array}%
\right) \left(
\begin{array}{c}
\lambda _{k}^{(1)} \\
\lambda _{k}^{(2)} \\
... \\
\lambda _{k}^{(n-1)}%
\end{array}%
\right)  \notag   \\
&=&\frac{1}{2}(\lambda _{k}^{(1)T},\lambda _{k}^{(2)T},...,\lambda
_{k}^{(n-1)T})\left(
\begin{array}{cccccc}
0 & \omega & \omega & ... & \omega & \omega \\
-\omega & 0 & \omega & ... & \omega & \omega \\
-\omega & -\omega & 0 & ... & \omega & \omega \\
... & ... & ... & ... & ... & ... \\
-\omega & -\omega & -\omega & ... & 0 & \omega \\
-\omega & -\omega & -\omega & ... & -\omega & 0%
\end{array}%
\right) \left(
\begin{array}{c}
\lambda _{k}^{(1)} \\
\lambda _{k}^{(2)} \\
... \\
\lambda _{k}^{(n-1)}%
\end{array}%
\right) .   \notag \label{eqA-16}
\end{eqnarray}

Expand $\lambda _{k}^{(j)T}\omega \lambda _{k}^{(j^{\prime })}$ as
\begin{equation}
\lambda _{k}^{(j)T}\omega \lambda _{k}^{(j^{\prime })}=\xi _{2k-1}^{(j)}\xi
_{2k}^{(j^{\prime })}-\xi _{2k}^{(j)}\xi _{2k-1}^{(j^{\prime })},    \notag \label{eqA-17}
\end{equation}
then in Eq. (\ref{eqA-12}), one finds
\begin{eqnarray}
\sum_{k=1}^{m}\sum_{1\leq j<j^{\prime }\leq n}\lambda _{k}^{(j)T}\omega
\lambda _{k}^{(j^{\prime })} &=&\sum_{1\leq j<j^{\prime }\leq
n-1}\sum_{k=1}^{m}(\xi _{2k-1}^{(j)}\xi _{2k}^{(j^{\prime })}-\xi
_{2k}^{(j)}\xi _{2k-1}^{(j^{\prime })})  \notag  \\
&=&\frac{1}{2}(\xi ^{(1)T},\xi ^{(2)T},...,\xi ^{(n-1)T})\left(
\begin{array}{cccccc}
0 & \Omega & \Omega & ... & \Omega & \Omega \\
-\Omega & 0 & \Omega & ... & \Omega & \Omega \\
-\Omega & -\Omega & 0 & ... & \Omega & \Omega \\
... & ... & ... & ... & ... & ... \\
-\Omega & -\Omega & -\Omega & ... & 0 & \Omega \\
-\Omega & -\Omega & -\Omega & ... & -\Omega & 0%
\end{array}%
\right) \left(
\begin{array}{c}
\xi ^{(1)} \\
\xi ^{(2)} \\
... \\
\xi ^{(n-1)}%
\end{array}%
\right) .    \label{eqA-18}
\end{eqnarray}

Now, inserting Eqs. (\ref{eqA-14},\ref{eqA-18}) into Eq. (\ref{eqA-12}), we get
\begin{eqnarray}
\text{tr}(\rho _{1}\rho _{2}...\rho _{n})
&=&\int \frac{d^{2m}\xi ^{(1)}}{\pi ^{m}}\int \frac{d^{2m}\xi ^{(2)}}{\pi
^{m}}...\int \frac{d^{2m}\xi ^{(n-1)}}{\pi ^{m}}\exp \left( -\frac{1}{2}%
\Gamma ^{T}M^{\prime }\Gamma +\Lambda ^{\prime T}\Gamma \right) ,   \label{eqA-19}  \\
\Gamma &=&(\xi ^{(1)T},\xi ^{(2)T},...,\xi ^{(n-1)T})^{T},    \notag \label{eqA-20}  \\
M' &=&\left(
\begin{array}{cccccc}
V^{(n)^{\prime }}+V^{(1)^{\prime }} & V^{(n)^{\prime }}+i\Omega &
V^{(n)^{\prime }}+i\Omega & ... & V^{(n)^{\prime }}+i\Omega & V^{(n)^{\prime
}}+i\Omega \\
V^{(n)^{\prime }}-i\Omega & V^{(n)^{\prime }}+V^{(2)^{\prime }} &
V^{(n)^{\prime }}-i\Omega & ... & V^{(n)^{\prime }}+i\Omega & V^{(n)^{\prime
}}+i\Omega \\
V^{(n)^{\prime }}-i\Omega & V^{(n)^{\prime }}-i\Omega & V^{(n)^{\prime
}}+V^{(3)^{\prime }} & ... & V^{(n)^{\prime }}+i\Omega & V^{(n)^{\prime
}}+i\Omega \\
... & ... & ... & ... & ... & ... \\
V^{(n)^{\prime }}-i\Omega & V^{(n)^{\prime }}-i\Omega & V^{(n)^{\prime
}}-i\Omega & ... & V^{(n)^{\prime }}+V^{(n-2)^{\prime }} & V^{(n)^{\prime
}}+i\Omega \\
V^{(n)^{\prime }}-i\Omega & V^{(n)^{\prime }}-i\Omega & V^{(n)^{\prime
}}-i\Omega & ... & V^{(n)^{\prime }}-i\Omega & V^{(n)^{\prime
}}+V^{(n-1)^{\prime }}%
\end{array}%
\right) ,   \label{eqA-21} \\
\Lambda ^{\prime } &=&-i([\Omega (\overline{X}^{(1)}-\overline{X}%
^{(n)})]^{T},[\Omega (\overline{X}^{(2)}-\overline{X}^{(n)})]^{T},...,[%
\Omega (\overline{X}^{(n-1)}-\overline{X}^{(n)})]^{T}),     \notag \label{eqA-22}
\end{eqnarray}
where $V^{(j)^{\prime }}=\Omega V^{(j)^{\prime }}\Omega ^{T}.$

Clearly, $M'=M'^{T},$ and
\begin{eqnarray*}
\text{Re}M' &=&\left(
\begin{array}{cccccc}
V^{(n)^{\prime }}+V^{(1)^{\prime }} & V^{(n)^{\prime }} & V^{(n)^{\prime }}
& ... & V^{(n)^{\prime }} & V^{(n)^{\prime }} \\
V^{(n)^{\prime }} & V^{(n)^{\prime }}+V^{(2)^{\prime }} & V^{(n)^{\prime }}
& ... & V^{(n)^{\prime }} & V^{(n)^{\prime }} \\
V^{(n)^{\prime }} & V^{(n)^{\prime }} & V^{(n)^{\prime }}+V^{(3)^{\prime }}
& ... & V^{(n)^{\prime }} & V^{(n)^{\prime }} \\
... & ... & ... & ... & ... & ... \\
V^{(n)^{\prime }} & V^{(n)^{\prime }} & V^{(n)^{\prime }} & ... &
V^{(n)^{\prime }}+V^{(n-2)^{\prime }} & V^{(n)^{\prime }} \\
V^{(n)^{\prime }} & V^{(n)^{\prime }} & V^{(n)^{\prime }} & ... &
V^{(n)^{\prime }} & V^{(n)^{\prime }}+V^{(n-1)^{\prime }}%
\end{array}%
\right) =P_{1}+P_{2}, \\
P_{1} &=&\left(
\begin{array}{cccccc}
V^{(1)^{\prime }} & 0 & 0 & 0 & 0 & 0 \\
0 & V^{(2)^{\prime }} & 0 & ... & 0 & 0 \\
0 & 0 & V^{(3)^{\prime }} & ... & 0 & 0 \\
... & ... & ... & ... & ... & ... \\
0 & 0 & 0 & ... & V^{(n-2)^{\prime }} & 0 \\
0 & 0 & 0 & ... & 0 & V^{(n-1)^{\prime }}%
\end{array}%
\right) ,\text{ }P_{2}=\left(
\begin{array}{cccccc}
V^{(n)^{\prime }} & V^{(n)^{\prime }} & V^{(n)^{\prime }} & ... &
V^{(n)^{\prime }} & V^{(n)^{\prime }} \\
V^{(n)^{\prime }} & V^{(n)^{\prime }} & V^{(n)^{\prime }} & ... &
V^{(n)^{\prime }} & V^{(n)^{\prime }} \\
V^{(n)^{\prime }} & V^{(n)^{\prime }} & V^{(n)^{\prime }} & ... &
V^{(n)^{\prime }} & V^{(n)^{\prime }} \\
... & ... & ... & ... & ... & ... \\
V^{(n)^{\prime }} & V^{(n)^{\prime }} & V^{(n)^{\prime }} & ... &
V^{(n)^{\prime }} & V^{(n)^{\prime }} \\
V^{(n)^{\prime }} & V^{(n)^{\prime }} & V^{(n)^{\prime }} & ... &
V^{(n)^{\prime }} & V^{(n)^{\prime }}%
\end{array}%
\right) .
\end{eqnarray*}
Since $V^{(j)}\succ 0$ and $\Omega ^{-1}=\Omega ^{T},$ then $V^{(j)^{\prime
}}=\Omega V^{(j)^{\prime }}\Omega ^{T}\succ 0,$ $P_{1}=\oplus
_{j=1}^{n-1}V^{(j)^{\prime }}\succ 0.$ $P_{2}$ can be recast as
\begin{equation*}
P_{2}=\left( \left(
\begin{array}{c}
1 \\
1 \\
... \\
1%
\end{array}%
\right) (1,1,...,1)\right) \otimes V^{(n)^{\prime }},
\end{equation*}
the tensor product of two positive semidefinite matrices, then $P_{2}\succeq
0.$ This shows $\text{Re}M'\succ 0.$

Applying Lemma \ref{Lemma-1} to Eq. (\ref{eqA-19}), one gets
\begin{eqnarray}
&&\text{tr}(\rho _{1}\rho _{2}...\rho _{n})=\frac{2^{m(n-1)}}{\sqrt{\det
M^{\prime }}}\exp \left( \frac{1}{2}\Lambda ^{\prime T}M^{\prime -1}\Lambda
^{\prime }\right).  \label{eqA-23}
\end{eqnarray}

Compare $M'$ and $M$ in Eqs. (\ref{eqA-21},\ref{eq3-2}), one finds
\begin{eqnarray}
M^{\prime }=\left(\oplus_{j=1}^{n-1}\Omega\right) M \left(\oplus_{j=1}^{n-1}\Omega^{T} \right), \ \ \ \
M^{\prime -1}=\left(\oplus_{j=1}^{n-1}\Omega^{T} \right) M^{-1} \left(\oplus_{j=1}^{n-1}\Omega \right),   \label{eqA-24}
\end{eqnarray}
and $\det M^{\prime }=\det M.$  Taking Eq. (\ref{eqA-24}) into Eq. (\ref{eqA-23}), Theorem \ref{Theorem-1} then follows.
\end{widetext}

\hypertarget{Appendix-B}{}
\section*{Appendix B: Proof that two expressions of tr$(\rho ^{n})$ in Corollary \ref{Corollary-1} and in Eq. (\ref{eq3-15}) are equal}
\setcounter{equation}{0} \renewcommand%
\theequation{B\arabic{equation}}

For $V,$ there exists a symplectic matrix $S$ such that Eqs. (\ref{eq3-9},\ref{eq3-10},\ref{eq3-11}) hold. We
rewrite $M_{V}$ in Eq. (\ref{eq3-6}) as
\begin{eqnarray}
M_{V} &=&(\oplus _{j=1}^{n-1}S)\overline{M_{V}}(\oplus _{j=1}^{n-1}S^{T}),   \notag \label{eqB-1}  \\
\overline{M_{V}} &=&\left(
\begin{array}{ccccc}
2V^{\oplus } & V^{\oplus }+i\Omega & V^{\oplus }+i\Omega & ... & V^{\oplus
}+i\Omega \\
V^{\oplus }-i\Omega & 2V^{\oplus } & V+i\Omega & ... & V^{\oplus }+i\Omega
\\
V^{\oplus }-i\Omega & V^{\oplus }-i\Omega & 2V^{\oplus } & ... & V^{\oplus
}+i\Omega \\
... & ... & ... & ... & ... \\
V^{\oplus }-i\Omega & V^{\oplus }-i\Omega & V^{\oplus }-i\Omega & ... &
2V^{\oplus }%
\end{array}%
\right) .  \notag  \label{eqB-2}
\end{eqnarray}

Any real symplectic matrix $S$ has the determinant $\det S=1$ \cite{Rump-2017-LAA}.
Consequently, $\det M_{V}=\det \overline{M_{V}}.$ Comparing the expressions of tr$%
(\rho ^{n})$ in Corollary \ref{Corollary-1} and in Eq. (\ref{eq3-15}), we only need to prove
\begin{equation}
\det \overline{M_{V}}=2^{2m(n-1)}\prod_{k=1}^{m}\left[ \frac{(\nu
_{k}+1)^{n}-(\nu _{k}-1)^{n}}{2^{n}}\right] ^{2}.  \label{eqB-3}
\end{equation}

We label the row (column) index of $\overline{M_{V}}$ by $%
(1,2,...,2m;2m+1,2m+2,...,2m+2m;...;2m(n-2)+1,2m(n-2)+2,...,2m(n-2)+2m).$ We
introduce the $2m(n-1)\times 2m(n-1)$ permutation matrix $P$ such that $P%
\overline{M_{V}}$ ($\overline{M_{V}}P^{T}$) is the matrix obtained by
reordering the rows (columns) of $\overline{M_{V}}$ as $%
(1,2,2m+1,2m+2,4m+1,4m+2,...,2m(n-2)+1,2m(n-2)+2;3,4,2m+3,2m+4,4m+3,4m+4,...,2m(n-2)+3,2m(n-2)+4;...;(2m-1),2m,2m+(2m-1),2m+2m,4m+(2m-1),4m+2m,...,2m(n-2)+(2m-1),2m(n-2)+2m).
$
We find that
\begin{eqnarray}
&&\det \overline{M_{V}}=\det (P\overline{M_{V}}P^{T}),   \notag \label{eqB-4} \\
&&P\overline{M_{V}}P^{T}=\oplus _{k=1}^{m}\overline{M_{k}},   \notag \label{eqB-5} \\
&&\overline{M_{k}}=\left(
\begin{array}{ccccc}
2\nu _{k}I_{2} & \nu _{k}I_{2}+i\omega & \nu _{k}I_{2}+i\omega & ... & \nu
_{k}I_{2}+i\omega \\
\nu _{k}I_{2}-i\omega & 2\nu _{k}I_{2} & \nu _{k}I_{2}+i\omega & ... & \nu
_{k}I_{2}+i\omega \\
\nu _{k}I_{2}-i\omega & \nu _{k}I_{2}-i\omega & 2\nu _{k}I_{2} & ... & \nu
_{k}I_{2}+i\omega \\
... & ... & ... & ... & ... \\
\nu _{k}I_{2}-i\omega & \nu _{k}I_{2}-i\omega & \nu _{k}I_{2}-i\omega & ...
& 2\nu _{k}I_{2}%
\end{array}%
\right), \ \ \ \      \notag \label{eqB-6} \\
&&\det \overline{M_{V}}=\prod_{k=1}^{m}\det \overline{M_{k}}.  \label{eqB-7}
\end{eqnarray}

Compare Eqs. (\ref{eqB-3},\ref{eqB-7}), we then only need to prove
\begin{equation}
\det \overline{M_{k}}=\left[ \frac{(\nu _{k}+1)^{n}-(\nu _{k}-1)^{n}}{2}%
\right] ^{2}   \label{eqB-8}
\end{equation}
for any $1\leq k\leq m.$

For $n=2,$ $\overline{M_{k}}=2\nu _{k}I_{2},$ one can directly check that Eq.
(\ref{eqB-8}) holds.

For $n=3,$
\begin{equation}
\overline{M_{k}}=\left(
\begin{array}{cc}
2\nu _{k}I_{2} & \nu _{k}I_{2}+i\omega \\
\nu _{k}I_{2}-i\omega & 2\nu _{k}I_{2}%
\end{array}%
\right) ,  \notag
\end{equation}
then one can directly check that $\det \overline{M_{V}}=(3\nu
_{k}^{2}+1)^{2},$  Eq. (\ref{eqB-8}) holds.

We consider the case of any $n\geq 2.$ Let%
\begin{eqnarray}
U=\frac{1}{\sqrt{2}}\left(
\begin{array}{cc}
1 & -i \\
1 & i%
\end{array}%
\right) , \ \
Z=\left(
\begin{array}{cc}
1 & 0 \\
0 & -1%
\end{array}%
\right) .    \notag \label{eqB-9}
\end{eqnarray}
$U$ is a unitary matrix, $Z$ is one of the Pauli matrices. We can check that
\begin{equation}
U\omega U^{\dagger }=iZ.    \notag \label{eqB-10}
\end{equation}
As a result,
\begin{eqnarray}
&&\overline{\overline{M_{k}}}=\left( \oplus _{k=1}^{n-1}U\right) \overline{%
M_{k}}\left( \oplus _{k=1}^{n-1}U^{\dagger }\right)  \notag \\
&&=\left(
\begin{array}{ccccc}
2\nu _{k}I_{2} & \nu _{k}I_{2}-Z & \nu _{k}I_{2}-Z & ... & \nu _{k}I_{2}-Z
\\
\nu _{k}I_{2}+Z & 2\nu _{k}I_{2} & \nu _{k}I_{2}-Z & ... & \nu _{k}I_{2}-Z
\\
\nu _{k}I_{2}+Z & \nu _{k}I_{2}+Z & 2\nu _{k}I_{2} & ... & \nu _{k}I_{2}-Z
\\
... & ... & ... & ... & ... \\
\nu _{k}I_{2}+Z & \nu _{k}I_{2}+Z & \nu _{k}I_{2}+Z & ... & 2\nu _{k}I_{2}%
\end{array}%
\right)  \notag \\
&&=\left(
\begin{array}{ccccc}
\begin{array}{cc}
2\nu _{k} & 0 \\
0 & 2\nu _{k}%
\end{array}
&
\begin{array}{cc}
\nu _{k}-1 & 0 \\
0 & \nu _{k}+1%
\end{array}
& ... &
\begin{array}{cc}
\nu _{k}-1 & 0 \\
0 & \nu _{k}+1%
\end{array}
\\
\begin{array}{cc}
\nu _{k}+1 & 0 \\
0 & \nu _{k}-1%
\end{array}
&
\begin{array}{cc}
2\nu _{k} & 0 \\
0 & 2\nu _{k}%
\end{array}
& ... &
\begin{array}{cc}
\nu _{k}-1 & 0 \\
0 & \nu _{k}+1%
\end{array}
\\
\begin{array}{cc}
\nu _{k}+1 & 0 \\
0 & \nu _{k}-1%
\end{array}
&
\begin{array}{cc}
\nu _{k}+1 & 0 \\
0 & \nu _{k}-1%
\end{array}
& ... &
\begin{array}{cc}
\nu _{k}-1 & 0 \\
0 & \nu _{k}+1%
\end{array}
\\
... & ... & ... &  ... \\
\begin{array}{cc}
\nu _{k}+1 & 0 \\
0 & \nu _{k}-1%
\end{array}
&
\begin{array}{cc}
\nu _{k}+1 & 0 \\
0 & \nu _{k}-1%
\end{array}
& ... &
\begin{array}{cc}
2\nu _{k} & 0 \\
0 & 2\nu _{k}%
\end{array}%
\end{array}%
\right) , \ \ \notag \\  \notag \label{eqB-11} \\
&&\det \overline{\overline{M_{k}}}=\det \overline{M_{k}}.   \notag \label{eqB-12}
\end{eqnarray}
Notice that, $\overline{\overline{M_{k}}}$ is not symmetric, while $%
\overline{M_{k}}$ is symmetric.

To compute $\det \overline{\overline{M_{k}}},$ we introduce the $%
2(n-1)\times 2(n-1)$ permutation matrix $P_{k}$ such that $P_{k}%
\overline{\overline{M_{k}}}$ ($\overline{\overline{M_{k}}}P_{k}^{T}$) is the
matrix obtained by reordering the rows (columns) $%
(1,2,3,4,5,...,2(n-1)-1,2(n-1))$ of $\overline{\overline{M_{k}}}$ to $%
(1,3,5,...,2(n-1)-1,2,4,6,...,2(n-1)).$ Thus

\begin{eqnarray}
P_{k}\overline{\overline{M_{k}}}P_{k}^{T} &=&N_{k}\oplus N_{k}^{T},  \notag \label{eqB-13}  \\
N_{k} &=&\left(
\begin{array}{ccccc}
2\nu _{k} & \nu _{k}-1 & \nu _{k}-1 & ... & \nu _{k}-1 \\
\nu _{k}+1 & 2\nu _{k} & \nu _{k}-1 & ... & \nu _{k}-1 \\
\nu _{k}+1 & \nu _{k}+1 & 2\nu _{k} & ... & \nu _{k}-1 \\
... & ... & ... & ... & ... \\
\nu _{k}+1 & \nu _{k}+1 & \nu _{k}+1 & ... & 2\nu _{k}%
\end{array}%
\right) .   \notag \label{eqB-14}
\end{eqnarray}

Since $\det \overline{\overline{M_{k}}}=\det \left( P_{k}\overline{\overline{%
M_{k}}}P_{k}^{T}\right) $ and $\det N_{k}=\det N_{k}^{T},$ with Eq. (\ref{eqB-8}), we
then only need to prove
\begin{equation}
\det N_{k}=\frac{(\nu _{k}+1)^{n}-(\nu _{k}-1)^{n}}{2}.  \label{eqB-15}
\end{equation}

Let%
\begin{eqnarray}
u_{k} &=&\frac{\nu _{k}+1}{\nu _{k}-1}>1,   \label{eqB-16} \\
E_{1} &=&\left(
\begin{array}{ccccccc}
1 & 0 & 0 & ... & 0 & 0 & 0 \\
-1 & 1 & 0 & ... & 0 & 0 & 0 \\
0 & -1 & 1 & ... & 0 & 0 & 0 \\
... & .. & .. & ... & ... & ... & ... \\
0 & 0 & 0 & ... & 1 & 0 & 0 \\
0 & 0 & 0 & ... & -1 & 1 & 0 \\
0 & 0 & 0 & ... & 0 & -1 & 1%
\end{array}%
\right)       \notag  \label{eqB-17} \\
E_{2} &=&\left(
\begin{array}{cccccc}
1 & 0 & 0 & ... & 0 & 0 \\
u_{k} & 1 & 0 & ... & 0 & 0 \\
u_{k}^{2} & u_{k} & 1 & ... & 0 & 0 \\
u_{k}^{3} & u_{k}^{2} & u_{k} & ... & 0 & 0 \\
... & ... & ... & ... & 1 & 0 \\
u_{k}^{n-2} & u_{k}^{n-3} & u_{k}^{n-4} & ... & u_{k} & 1%
\end{array}%
\right) .    \notag  \label{eqB-18}
\end{eqnarray}
Clearly, $\det E_{1}=\det E_{2}=1,$ hence $\det N_{k}=\det
(E_{2}N_{k}E_{1}). $ We directly compute that $E_{2}(N_{k}E_{1})$ equals
\begin{equation}
\left(
\begin{array}{cccccc}
\nu _{k}-1 & 0 & 0 & ... & 0 & (\nu _{k}+1) \\
0 & \nu _{k}-1 & 0 & ... & 0 & (\nu _{k}+1)(1+u_{k}) \\
0 & 0 & \nu _{k}-1 & ... & 0 & (\nu _{k}+1)(1+u_{k}+u_{k}^{2}) \\
... & ... & ... & ... & ... & ... \\
0 & 0 & 0 & ... & \nu _{k}-1 & (\nu _{k}+1)\sum_{j=0}^{n-3}u_{k}^{j} \\
0 & 0 & 0 & ... & 0 & 2\nu _{k}+(\nu _{k}+1)\sum_{j=1}^{n-2}u_{k}^{j}%
\end{array}%
\right) .     \notag  \label{eqB-19}
\end{equation}
Consequently,
\begin{equation}
\det N_{k}=(\nu _{k}-1)^{n-2}\left[ 2\nu _{k}+(\nu
_{k}+1)\sum_{j=1}^{n-2}u_{k}^{j}\right] .    \notag  \label{eqB-20}
\end{equation}

Employing the sum of finite geometric series
\begin{equation}
\sum_{j=1}^{n-2}u_{k}^{j}=%
\frac{u_{k}(1-u_{k}^{n-2})}{1-u_{k}}=\frac{u_{k}(u_{k}^{n-2}-1)}{u_{k}-1} \notag   \label{eqB-21}
\end{equation}
and Eq. (\ref{eqB-16}), after some algebras, we will prove Eq. (\ref{eqB-15}). We then proved that
the expressions of tr$(\rho ^{n})$ in Corollary \ref{Corollary-1} and in Eq. (\ref{eq3-15}) are equal.

\hypertarget{Appendix-C}{}
\section*{Appendix C: Proof of Eq. (\ref{eq4-19})}
\setcounter{equation}{0} \renewcommand%
\theequation{C\arabic{equation}}

We first recall the identities
\begin{eqnarray}
\sum_{j=1}^{n}e^{i\frac{2\pi kj}{n}} &=&n\delta _{n,k},    \notag \label{eqC-1} \\
\sum_{j=1}^{n}\cos \frac{2\pi kj}{n} &=&n\delta _{n,k},   \notag \label{eqC-2} \\
\sum_{j=1}^{n}\sin \frac{2\pi kj}{n} &=&0,    \notag \label{eqC-3}
\end{eqnarray}
where $\delta _{n,k}=1$ if $k=0$ (mod $n),$ $\delta _{n,k}=0$  if $k\neq0$ (mod $n$).

In Eq. (\ref{eq4-3}), notice that the entries of $V^{-1}$ are
\begin{equation}
\begin{cases}
V_{11}^{-1}=\cosh (2|\zeta |)-\cos \phi \sinh (2|\zeta |), \\
V_{12}^{-1}=V_{21}^{-1}=-\sin \phi \sinh (2|\zeta |), \\
V_{22}^{-1}=\cosh (2|\zeta |)+\cos \phi \sinh (2|\zeta |),%
\end{cases}   \label{eqC-4}
\end{equation}
and
\begin{equation}
(V^{-1}+i\omega )^{T}=(V^{-1}-i\omega ).    \notag \label{eqC-5}
\end{equation}
Then to compute $\text{tr}\left( |\Phi _{1}\rangle \left\langle \Phi _{1}\right\vert \Phi
_{2}\rangle \left\langle \Phi _{2}\right\vert ...|\Phi _{n}\rangle
\left\langle \Phi _{n}\right\vert \right)$ by Theorem \ref{Theorem-1} with Eqs. (\ref{eq4-10},\ref{eq4-12},\ref{eq4-13},\ref{eqC-4}), we only need to compute the following two expressions $E_{1}$ and $E_{2}.$ We have
\begin{eqnarray}
E_{1} &=&\sum_{j=1}^{n}\left( \cos \frac{2\pi j}{n}-1,\sin \frac{2\pi j}{n}%
\right) V^{-1}\left( \cos \frac{2\pi j}{n}-1,\sin \frac{2\pi j}{n}\right)
^{T}  \notag  \\
&=&V_{11}^{-1}\sum_{j=1}^{n}\left( \cos \frac{2\pi j}{n}-1\right)
^{2}+V_{22}^{-1}\sum_{j=1}^{n}\sin ^{2}\frac{2\pi j}{n} \notag \\
&&+2V_{12}^{-1}%
\sum_{j=1}^{n}\left( \cos \frac{2\pi j}{n}-1\right) \sin \frac{2\pi j}{n}  \notag \\
&=&\frac{3n}{2}V_{11}^{-1}+\frac{n}{2}V_{22}^{-1}  \notag \\
&=&n[2\cosh (2|\zeta |)-\cos \phi \sinh (2|\zeta |)].   \label{eqC-6}
\end{eqnarray}%
and
\begin{eqnarray}
E_{2} &=&\sum_{j=1}^{n}\left( \cos \frac{2\pi j}{n}-1,\sin \frac{2\pi j}{n}%
\right) (V^{-1}+i\omega ) \notag \\
&& \ \ \ \  \cdot\left( \cos \frac{2\pi (j+1)}{n}-1,\sin \frac{2\pi
(j+1)}{n}\right) ^{T}  \notag  \\
&=&V_{11}^{-1}\sum_{j=1}^{n}\left( \cos \frac{2\pi j}{n}-1\right) \left(
\cos \frac{2\pi (j+1)}{n}-1\right)  \notag \\
&&+V_{22}^{-1}\sum_{j=1}^{n}\sin \frac{2\pi
j}{n}\sin \frac{2\pi (j+1)}{n} \notag \\
&&+(V_{12}^{-1}+i)\sum_{j=1}^{n}\left( \cos \frac{2\pi j}{n}-1\right) \sin
\frac{2\pi (j+1)}{n}   \notag  \\
&&+(V_{12}^{-1}-i)\sum_{j=1}^{n}\sin \frac{2\pi j}{n}%
\left( \cos \frac{2\pi (j+1)}{n}-1\right) \notag \\
&=&\frac{n}{2}\left( 2+\cos \frac{2\pi }{n}\right) V_{11}^{-1}+\frac{n}{2}%
V_{22}^{-1}\cos \frac{2\pi }{n}   \notag  \\
&&+\frac{n}{2}(V_{12}^{-1}+i)\sin \frac{2\pi }{n%
}-\frac{n}{2}(V_{12}^{-1}-i)\sin \frac{2\pi }{n} \notag \\
&=&n\left[ \left( 1+\cos \frac{2\pi }{n}\right) \cosh (2|\zeta |)-\cos \phi
\sinh (2|\zeta |)+i\sin \frac{2\pi }{n}\right]. \notag \\ \label{eqC-7}
\end{eqnarray}
In the above derivations of $E_{1}$ and $E_{2},$ we have used many common
trigonometric formulas, such as double angle formulas, product-to-sum
formulas, angle sum and difference formulas. Notice that since $\left( \cos
\frac{2\pi j}{n}-1,\sin \frac{2\pi j}{n}\right) =\left( 0,0\right) $ for $%
j=n,$ then $\sum_{j=1}^{n}=\sum_{j=1}^{n-1}$ in $E_{1},$ $%
\sum_{j=1}^{n}=\sum_{j=1}^{n-2}$ in $E_{2}.$

Taking Eqs. (\ref{eq4-10},\ref{eq4-12},\ref{eq4-13},\ref{eqC-4},\ref{eqC-6},\ref{eqC-7}) into Eq. (\ref{eq3-1}), we get
\begin{eqnarray}
&&\text{tr}\left( |\Phi _{1}\rangle \left\langle \Phi _{1}\right\vert \Phi
_{2}\rangle \left\langle \Phi _{2}\right\vert ...|\Phi _{n}\rangle
\left\langle \Phi _{n}\right\vert \right) \notag \\
&=&\exp \left[ -\frac{1}{2}\Lambda ^{T}M_{Vp}^{-1}\Lambda \right] \notag \\
&=&\exp \left[ -|\alpha |^{2}(E_{1}-E_{2})\right] \notag \\
&=&\exp \left\{ -n|\alpha |^{2}\left[ \left( 1-\cos \frac{2\pi }{n}\right)
\cosh (2|\zeta |)-i\sin \frac{2\pi }{n}\right] \right\} \notag \\
&=&\exp \left\{ -2n|\alpha |^{2}\sin ^{2}\frac{\pi }{n}\left[ \cosh (2|\zeta
|)-i\cot \frac{\pi }{n}\right] \right\} .    \notag \label{eqC-8}
\end{eqnarray}
This completes the proof of Eq. (\ref{eq4-19}).

\hypertarget{Appendix-D}{}
\section*{Appendix D: Proof of Eq. (\ref{eq4-41})}
\setcounter{equation}{0} \renewcommand%
\theequation{D\arabic{equation}}

Let
\begin{eqnarray}
u &=&\sinh ^{2}|\zeta |\geq 0,   \notag \label{eqD-1} \\
se^{i\beta } &=&\cosh ^{2}|\zeta |-e^{i\frac{2\pi }{n}}\sinh ^{2}|\zeta |,   \notag \label{eqD-2}
\end{eqnarray}%
with $s>0$, $\beta \in\mathbb{R}.$ Thus
\begin{eqnarray}
se^{i\beta }&=&\cosh ^{2}|\zeta |-e^{i\frac{2\pi }{n}}\sinh ^{2}|\zeta |   \notag \\
&=&1+u\left( 1-\cos \frac{2\pi }{n}\right) -iu\sin \frac{2\pi }{n},  \label{eqD-3} \\
se^{i\beta }&=&s\cos \beta +is\sin \beta . \label{eqD-4}
\end{eqnarray}

Comparing Eqs. (\ref{eqD-3},\ref{eqD-4}), we find
\begin{eqnarray}
s\cos \beta &=&1+u\left( 1-\cos \frac{2\pi }{n}\right) \geq 1,   \label{eqD-5}  \\
s\sin \beta &=&-u\sin \frac{2\pi }{n}\in (-\infty,0],    \label{eqD-6}   \\
\tan \beta &=&-\frac{u\sin \frac{2\pi }{n}}{1+u\left( 1-\cos \frac{2\pi }{n}%
\right) }\in (-\cot \frac{\pi }{n},0],  \label{eqD-7}
\end{eqnarray}
in Eq. (\ref{eqD-7}), we have used the trigonometric identity
\begin{equation}
\frac{\sin \frac{2\pi }{n}}{1-\cos \frac{2\pi }{n}}=\frac{1+\cos \frac{2\pi
}{n}}{\sin \frac{2\pi }{n}}=\cot \frac{\pi }{n}.  \notag
\end{equation}

Based on Eqs. (\ref{eqD-5},\ref{eqD-6},\ref{eqD-7}), we let
\begin{equation}
\beta \in \left( \frac{\pi }{n}-\frac{\pi }{2},0\right] .    \notag \label{eqD-8}
\end{equation}
From Eq. (\ref{eqD-6}), one has
\begin{equation}
u=-\frac{s\sin \beta }{\sin \frac{2\pi }{n}}.   \label{eqD-9}
\end{equation}

Plug Eq. (\ref{eqD-9}) into Eq. (\ref{eqD-5}), one gets
\begin{eqnarray}
&& s\cos \beta =1-s\sin \beta \tan \frac{\pi }{n},   \notag  \\
&& s\left( \cos \beta +\sin \beta \tan \frac{\pi }{n}\right) =1,  \notag  \\
&& s\left( \cos \beta \cos \frac{\pi }{n}+\sin \beta \sin \frac{\pi }{n}\right)=\cos \frac{\pi }{n},   \notag   \\
&& s=\frac{\cos \frac{\pi }{n}}{\cos \left( \beta -\frac{\pi }{n}\right) }.   \label{eqD-10}
\end{eqnarray}

Let
\begin{equation}
re^{i\theta }=\left( \cosh ^{2}|\zeta |-e^{i\frac{2\pi }{n}}\sinh ^{2}|\zeta
|\right) ^{-\frac{n}{2}},     \notag \label{eqD-11}
\end{equation}%
with $r>0$, $\theta \in\mathbb{R}.$ Combining Eq. (\ref{eqD-10}), we get
\begin{eqnarray}
re^{i\theta } &=&\left( se^{i\beta }\right) ^{-\frac{n}{2}}=\left( \frac{%
\cos \left( \beta -\frac{\pi }{n}\right) }{\cos \frac{\pi }{n}}\right) ^{%
\frac{n}{2}}e^{-i\frac{n}{2}\beta },  \notag \\
\theta  &=&-\frac{n}{2}\beta \in \left[ 0,\frac{n-2}{4}\pi \right) ,  \notag \\
r &=&\left( \cos \frac{2\theta +\pi }{n}\sec \frac{\pi }{n}\right) ^{\frac{n%
}{2}}.  \notag
\end{eqnarray}
We then proved Eq. (\ref{eq4-41}).
\\

\hypertarget{Appendix-E}{}
\section*{Appendix E: Proof of Eqs. (\ref{eq4-31},\ref{eq4-42})}
\setcounter{equation}{0} \renewcommand%
\theequation{E\arabic{equation}}

(E.a). Proof of Eq. (\ref{eq4-31}).

For $\theta \in \lbrack 0,\pi ],$ since both sides of Eq. (\ref{eq4-31}) are positive,
then take their positive $n$th roots,
inequality (\ref{eq4-31}) is equivalent to
\begin{equation}
e^{-\theta \frac{\tan \frac{\pi }{n}}{n}}\leq \cos \frac{\pi }{n}\sec \frac{%
\pi -\theta }{n},  \notag
\end{equation}
and further is equivalent to
\begin{equation}
\cos \frac{\pi -\theta }{n}\leq \cos \frac{\pi }{n}e^{\theta \frac{\tan
\frac{\pi }{n}}{n}}.   \label{eqE-1}
\end{equation}

Since both sides of Eq. (\ref{eqE-1}) are positive, then take their natural logarithms,
inequality (\ref{eqE-1}) is equivalent to
\begin{equation}
\ln \cos \frac{\pi -\theta }{n}\leq \ln \cos \frac{\pi }{n}+\theta \frac{%
\tan \frac{\pi }{n}}{n}.   \notag
\end{equation}

Define the function
\begin{equation}
f(\theta )=\ln \cos \frac{\pi }{n}+\theta \frac{\tan \frac{\pi }{n}}{n}-\ln
\cos \frac{\pi -\theta }{n},   \notag
\end{equation}
we need to prove $f(\theta )\geq 0$ for $\theta \in \lbrack 0,\pi ].$

Taking the derivative of $f(\theta ),$ one has
\begin{equation}
\frac{d}{d\theta }f(\theta )=\frac{\tan \frac{\pi }{n}}{n}-\frac{\tan \frac{%
\pi -\theta }{n}}{n}\geq 0   \notag
\end{equation}
with equality if and only if $\theta =0.$ With $f(0)=0,$ we then proved that
inequality (\ref{eq4-31}) holds.

(E.b). Proof of Eq. (\ref{eq4-42}).

For $\theta \in \left[ 0,\min \{\pi ,\frac{n-2}{4}\pi \}\right] ,$ notice that
both sides of Eq. (\ref{eq4-42}) are positive, then take their positive $\frac{n}{2}$th
roots, one gets the equivalent inequality
\begin{equation}
\cos \frac{2\theta +\pi }{n}\sec \frac{\pi }{n}\leq e^{-\frac{2}{n}\theta
\tan \frac{\pi }{n}}. \label{eqE-2}
\end{equation}

Since both sides of Eq. (\ref{eqE-2}) are positive, then take their natural logarithms,
we further get the equivalent inequality
\begin{equation}
\ln \cos \frac{2\theta +\pi }{n}+\ln \sec \frac{\pi }{n}+\frac{2}{n}\theta
\tan \frac{\pi }{n}\leq 0. \notag
\end{equation}

Define the function
\begin{equation}
h(\theta )=\ln \cos \frac{2\theta +\pi }{n}+\ln \sec \frac{\pi }{n}+\frac{2}{%
n}\theta \tan \frac{\pi }{n}, \notag
\end{equation}
we now prove $h(\theta )\leq 0$ for $\theta \in \left[ 0,\min \{\pi ,\frac{%
n-2}{4}\pi \}\right] .$

The derivative of $h(\theta )$ is
\begin{equation}
\frac{d}{d\theta }h(\theta )=\frac{2}{n}\left( \tan \frac{\pi }{n}-\tan
\frac{2\theta +\pi }{n}\right) \leq 0 \notag
\end{equation}
with equality if and only if $\theta =0.$ With $h(0)=0,$ we then proved
inequality (\ref{eq4-42}).


\begin{thebibliography}{61}%
\makeatletter
\providecommand \@ifxundefined [1]{%
 \@ifx{#1\undefined}
}%
\providecommand \@ifnum [1]{%
 \ifnum #1\expandafter \@firstoftwo
 \else \expandafter \@secondoftwo
 \fi
}%
\providecommand \@ifx [1]{%
 \ifx #1\expandafter \@firstoftwo
 \else \expandafter \@secondoftwo
 \fi
}%
\providecommand \natexlab [1]{#1}%
\providecommand \enquote  [1]{``#1''}%
\providecommand \bibnamefont  [1]{#1}%
\providecommand \bibfnamefont [1]{#1}%
\providecommand \citenamefont [1]{#1}%
\providecommand \href@noop [0]{\@secondoftwo}%
\providecommand \href [0]{\begingroup \@sanitize@url \@href}%
\providecommand \@href[1]{\@@startlink{#1}\@@href}%
\providecommand \@@href[1]{\endgroup#1\@@endlink}%
\providecommand \@sanitize@url [0]{\catcode `\\12\catcode `\$12\catcode
  `\&12\catcode `\#12\catcode `\^12\catcode `\_12\catcode `\%12\relax}%
\providecommand \@@startlink[1]{}%
\providecommand \@@endlink[0]{}%
\providecommand \url  [0]{\begingroup\@sanitize@url \@url }%
\providecommand \@url [1]{\endgroup\@href {#1}{\urlprefix }}%
\providecommand \urlprefix  [0]{URL }%
\providecommand \Eprint [0]{\href }%
\providecommand \doibase [0]{http://dx.doi.org/}%
\providecommand \selectlanguage [0]{\@gobble}%
\providecommand \bibinfo  [0]{\@secondoftwo}%
\providecommand \bibfield  [0]{\@secondoftwo}%
\providecommand \translation [1]{[#1]}%
\providecommand \BibitemOpen [0]{}%
\providecommand \bibitemStop [0]{}%
\providecommand \bibitemNoStop [0]{.\EOS\space}%
\providecommand \EOS [0]{\spacefactor3000\relax}%
\providecommand \BibitemShut  [1]{\csname bibitem#1\endcsname}%
\let\auto@bib@innerbib\@empty
\bibitem [{\citenamefont {Bargmann}(1964)}]{Bargmann-1964-JMP}%
  \BibitemOpen
  \bibfield  {author} {\bibinfo {author} {\bibfnamefont {V.}~\bibnamefont
  {Bargmann}},\ }\bibinfo {title} {Note on wigner's theorem on symmetry
  operations},\ \href {\doibase 10.1063/1.1704188} {\bibfield  {journal}
  {\bibinfo  {journal} {Journal of Mathematical Physics}\ }\textbf {\bibinfo
  {volume} {5}},\ \bibinfo {pages} {862} (\bibinfo {year} {1964})}\BibitemShut
  {NoStop}%
\bibitem [{\citenamefont {Quek}\ \emph {et~al.}(2024)\citenamefont {Quek},
  \citenamefont {Kaur},\ and\ \citenamefont {Wilde}}]{Wilde-2024-Quantum}%
  \BibitemOpen
  \bibfield  {author} {\bibinfo {author} {\bibfnamefont {Y.}~\bibnamefont
  {Quek}}, \bibinfo {author} {\bibfnamefont {E.}~\bibnamefont {Kaur}}, \ and\
  \bibinfo {author} {\bibfnamefont {M.~M.}\ \bibnamefont {Wilde}},\ }\bibinfo
  {title} {Multivariate trace estimation in constant quantum depth},\ \href
  {\doibase 10.22331/q-2024-01-10-1220} {\bibfield  {journal} {\bibinfo
  {journal} {{Quantum}}\ }\textbf {\bibinfo {volume} {8}},\ \bibinfo {pages}
  {1220} (\bibinfo {year} {2024})}\BibitemShut {NoStop}%
\bibitem [{\citenamefont {Mukunda}\ \emph {et~al.}(2001)\citenamefont
  {Mukunda}, \citenamefont {Arvind}, \citenamefont {Chaturvedi},\ and\
  \citenamefont {Simon}}]{Simon-2001-PRA}%
  \BibitemOpen
  \bibfield  {author} {\bibinfo {author} {\bibfnamefont {N.}~\bibnamefont
  {Mukunda}}, \bibinfo {author} {\bibnamefont {Arvind}}, \bibinfo {author}
  {\bibfnamefont {S.}~\bibnamefont {Chaturvedi}}, \ and\ \bibinfo {author}
  {\bibfnamefont {R.}~\bibnamefont {Simon}},\ }\bibinfo {title} {Bargmann
  invariants and off-diagonal geometric phases for multilevel quantum systems:
  A unitary-group approach},\ \href {\doibase 10.1103/PhysRevA.65.012102}
  {\bibfield  {journal} {\bibinfo  {journal} {Phys. Rev. A}\ }\textbf {\bibinfo
  {volume} {65}},\ \bibinfo {pages} {012102} (\bibinfo {year}
  {2001})}\BibitemShut {NoStop}%
\bibitem [{\citenamefont {Mukunda}\ \emph
  {et~al.}(2003{\natexlab{a}})\citenamefont {Mukunda}, \citenamefont {Arvind},
  \citenamefont {Ercolessi}, \citenamefont {Marmo}, \citenamefont {Morandi},\
  and\ \citenamefont {Simon}}]{Simon-2003-PRA}%
  \BibitemOpen
  \bibfield  {author} {\bibinfo {author} {\bibfnamefont {N.}~\bibnamefont
  {Mukunda}}, \bibinfo {author} {\bibnamefont {Arvind}}, \bibinfo {author}
  {\bibfnamefont {E.}~\bibnamefont {Ercolessi}}, \bibinfo {author}
  {\bibfnamefont {G.}~\bibnamefont {Marmo}}, \bibinfo {author} {\bibfnamefont
  {G.}~\bibnamefont {Morandi}}, \ and\ \bibinfo {author} {\bibfnamefont
  {R.}~\bibnamefont {Simon}},\ }\bibinfo {title} {Bargmann invariants, null
  phase curves, and a theory of the geometric phase},\ \href {\doibase
  10.1103/PhysRevA.67.042114} {\bibfield  {journal} {\bibinfo  {journal} {Phys.
  Rev. A}\ }\textbf {\bibinfo {volume} {67}},\ \bibinfo {pages} {042114}
  (\bibinfo {year} {2003}{\natexlab{a}})}\BibitemShut {NoStop}%
\bibitem [{\citenamefont {Mukunda}\ \emph
  {et~al.}(2003{\natexlab{b}})\citenamefont {Mukunda}, \citenamefont
  {Aravind},\ and\ \citenamefont {Simon}}]{Simon-2003-JPA}%
  \BibitemOpen
  \bibfield  {author} {\bibinfo {author} {\bibfnamefont {N.}~\bibnamefont
  {Mukunda}}, \bibinfo {author} {\bibfnamefont {P.~K.}\ \bibnamefont
  {Aravind}}, \ and\ \bibinfo {author} {\bibfnamefont {R.}~\bibnamefont
  {Simon}},\ }\bibinfo {title} {Wigner rotations, bargmann invariants and
  geometric phases},\ \href {\doibase 10.1088/0305-4470/36/9/312} {\bibfield
  {journal} {\bibinfo  {journal} {Journal of Physics A: Mathematical and
  General}\ }\textbf {\bibinfo {volume} {36}},\ \bibinfo {pages} {2347}
  (\bibinfo {year} {2003}{\natexlab{b}})}\BibitemShut {NoStop}%
\bibitem [{\citenamefont {Avdoshkin}\ and\ \citenamefont
  {Popov}(2023)}]{Popov-2023-PRB}%
  \BibitemOpen
  \bibfield  {author} {\bibinfo {author} {\bibfnamefont {A.}~\bibnamefont
  {Avdoshkin}}\ and\ \bibinfo {author} {\bibfnamefont {F.~K.}\ \bibnamefont
  {Popov}},\ }\bibinfo {title} {Extrinsic geometry of quantum states},\ \href
  {\doibase 10.1103/PhysRevB.107.245136} {\bibfield  {journal} {\bibinfo
  {journal} {Phys. Rev. B}\ }\textbf {\bibinfo {volume} {107}},\ \bibinfo
  {pages} {245136} (\bibinfo {year} {2023})}\BibitemShut {NoStop}%
\bibitem [{\citenamefont {Menssen}\ \emph {et~al.}(2017)\citenamefont
  {Menssen}, \citenamefont {Jones}, \citenamefont {Metcalf}, \citenamefont
  {Tichy}, \citenamefont {Barz}, \citenamefont {Kolthammer},\ and\
  \citenamefont {Walmsley}}]{Menssen-2017-PRL}%
  \BibitemOpen
  \bibfield  {author} {\bibinfo {author} {\bibfnamefont {A.~J.}\ \bibnamefont
  {Menssen}}, \bibinfo {author} {\bibfnamefont {A.~E.}\ \bibnamefont {Jones}},
  \bibinfo {author} {\bibfnamefont {B.~J.}\ \bibnamefont {Metcalf}}, \bibinfo
  {author} {\bibfnamefont {M.~C.}\ \bibnamefont {Tichy}}, \bibinfo {author}
  {\bibfnamefont {S.}~\bibnamefont {Barz}}, \bibinfo {author} {\bibfnamefont
  {W.~S.}\ \bibnamefont {Kolthammer}}, \ and\ \bibinfo {author} {\bibfnamefont
  {I.~A.}\ \bibnamefont {Walmsley}},\ }\bibinfo {title} {Distinguishability and
  many-particle interference},\ \href {\doibase 10.1103/PhysRevLett.118.153603}
  {\bibfield  {journal} {\bibinfo  {journal} {Phys. Rev. Lett.}\ }\textbf
  {\bibinfo {volume} {118}},\ \bibinfo {pages} {153603} (\bibinfo {year}
  {2017})}\BibitemShut {NoStop}%
\bibitem [{\citenamefont {Jones}\ \emph {et~al.}(2020)\citenamefont {Jones},
  \citenamefont {Menssen}, \citenamefont {Chrzanowski}, \citenamefont
  {Wolterink}, \citenamefont {Shchesnovich},\ and\ \citenamefont
  {Walmsley}}]{Jones-2020-PRL}%
  \BibitemOpen
  \bibfield  {author} {\bibinfo {author} {\bibfnamefont {A.~E.}\ \bibnamefont
  {Jones}}, \bibinfo {author} {\bibfnamefont {A.~J.}\ \bibnamefont {Menssen}},
  \bibinfo {author} {\bibfnamefont {H.~M.}\ \bibnamefont {Chrzanowski}},
  \bibinfo {author} {\bibfnamefont {T.~A.~W.}\ \bibnamefont {Wolterink}},
  \bibinfo {author} {\bibfnamefont {V.~S.}\ \bibnamefont {Shchesnovich}}, \
  and\ \bibinfo {author} {\bibfnamefont {I.~A.}\ \bibnamefont {Walmsley}},\
  }\bibinfo {title} {Multiparticle interference of pairwise distinguishable
  photons},\ \href {\doibase 10.1103/PhysRevLett.125.123603} {\bibfield
  {journal} {\bibinfo  {journal} {Phys. Rev. Lett.}\ }\textbf {\bibinfo
  {volume} {125}},\ \bibinfo {pages} {123603} (\bibinfo {year}
  {2020})}\BibitemShut {NoStop}%
\bibitem [{\citenamefont {Liang}\ \emph {et~al.}(2023)\citenamefont {Liang},
  \citenamefont {Lv}, \citenamefont {Wang},\ and\ \citenamefont
  {Fei}}]{Fei-2023-PRA}%
  \BibitemOpen
  \bibfield  {author} {\bibinfo {author} {\bibfnamefont {J.-M.}\ \bibnamefont
  {Liang}}, \bibinfo {author} {\bibfnamefont {Q.-Q.}\ \bibnamefont {Lv}},
  \bibinfo {author} {\bibfnamefont {Z.-X.}\ \bibnamefont {Wang}}, \ and\
  \bibinfo {author} {\bibfnamefont {S.-M.}\ \bibnamefont {Fei}},\ }\bibinfo
  {title} {Unified multivariate trace estimation and quantum error
  mitigation},\ \href {\doibase 10.1103/PhysRevA.107.012606} {\bibfield
  {journal} {\bibinfo  {journal} {Phys. Rev. A}\ }\textbf {\bibinfo {volume}
  {107}},\ \bibinfo {pages} {012606} (\bibinfo {year} {2023})}\BibitemShut
  {NoStop}%
\bibitem [{\citenamefont {Reascos}\ \emph {et~al.}(2023)\citenamefont
  {Reascos}, \citenamefont {Murta}, \citenamefont {Galv\~ao},\ and\
  \citenamefont {Fern\'andez-Rossier}}]{Galvao-2023-PRR}%
  \BibitemOpen
  \bibfield  {author} {\bibinfo {author} {\bibfnamefont {L.~I.}\ \bibnamefont
  {Reascos}}, \bibinfo {author} {\bibfnamefont {B.}~\bibnamefont {Murta}},
  \bibinfo {author} {\bibfnamefont {E.~F.}\ \bibnamefont {Galv\~ao}}, \ and\
  \bibinfo {author} {\bibfnamefont {J.}~\bibnamefont {Fern\'andez-Rossier}},\
  }\bibinfo {title} {Quantum circuits to measure scalar spin chirality},\ \href
  {\doibase 10.1103/PhysRevResearch.5.043087} {\bibfield  {journal} {\bibinfo
  {journal} {Phys. Rev. Res.}\ }\textbf {\bibinfo {volume} {5}},\ \bibinfo
  {pages} {043087} (\bibinfo {year} {2023})}\BibitemShut {NoStop}%
\bibitem [{\citenamefont {Kirkwood}(1933)}]{Kirkwood-1933-PR}%
  \BibitemOpen
  \bibfield  {author} {\bibinfo {author} {\bibfnamefont {J.~G.}\ \bibnamefont
  {Kirkwood}},\ }\bibinfo {title} {Quantum statistics of almost classical
  assemblies},\ \href {\doibase 10.1103/PhysRev.44.31} {\bibfield  {journal}
  {\bibinfo  {journal} {Phys. Rev.}\ }\textbf {\bibinfo {volume} {44}},\
  \bibinfo {pages} {31} (\bibinfo {year} {1933})}\BibitemShut {NoStop}%
\bibitem [{\citenamefont {Dirac}(1945)}]{Dirac-1945-RMP}%
  \BibitemOpen
  \bibfield  {author} {\bibinfo {author} {\bibfnamefont {P.~A.~M.}\
  \bibnamefont {Dirac}},\ }\bibinfo {title} {On the analogy between classical
  and quantum mechanics},\ \href {\doibase 10.1103/RevModPhys.17.195}
  {\bibfield  {journal} {\bibinfo  {journal} {Rev. Mod. Phys.}\ }\textbf
  {\bibinfo {volume} {17}},\ \bibinfo {pages} {195} (\bibinfo {year}
  {1945})}\BibitemShut {NoStop}%
\bibitem [{\citenamefont {Arvidsson-Shukur}\ \emph {et~al.}(2021)\citenamefont
  {Arvidsson-Shukur}, \citenamefont {Drori},\ and\ \citenamefont
  {Halpern}}]{NYH-2021-JPA}%
  \BibitemOpen
  \bibfield  {author} {\bibinfo {author} {\bibfnamefont {D.~R.~M.}\
  \bibnamefont {Arvidsson-Shukur}}, \bibinfo {author} {\bibfnamefont {J.~C.}\
  \bibnamefont {Drori}}, \ and\ \bibinfo {author} {\bibfnamefont {N.~Y.}\
  \bibnamefont {Halpern}},\ }\bibinfo {title} {Conditions tighter than
  noncommutation needed for nonclassicality},\ \href {\doibase
  10.1088/1751-8121/ac0289} {\bibfield  {journal} {\bibinfo  {journal} {J.
  Phys. A: Math. Theor.}\ }\textbf {\bibinfo {volume} {54}},\ \bibinfo {pages}
  {284001} (\bibinfo {year} {2021})}\BibitemShut {NoStop}%
\bibitem [{\citenamefont {Arvidsson-Shukur}\ \emph {et~al.}(2024)\citenamefont
  {Arvidsson-Shukur}, \citenamefont {Braasch~Jr}, \citenamefont {De~Bi\`evre},
  \citenamefont {Dressel}, \citenamefont {Jordan}, \citenamefont {Langrenez},
  \citenamefont {Lostaglio}, \citenamefont {Lundeen},\ and\ \citenamefont
  {Halpern}}]{Bievre-2024-NJP}%
  \BibitemOpen
  \bibfield  {author} {\bibinfo {author} {\bibfnamefont {D.~R.}\ \bibnamefont
  {Arvidsson-Shukur}}, \bibinfo {author} {\bibfnamefont {W.~F.}\ \bibnamefont
  {Braasch~Jr}}, \bibinfo {author} {\bibfnamefont {S.}~\bibnamefont
  {De~Bi\`evre}}, \bibinfo {author} {\bibfnamefont {J.}~\bibnamefont
  {Dressel}}, \bibinfo {author} {\bibfnamefont {A.~N.}\ \bibnamefont {Jordan}},
  \bibinfo {author} {\bibfnamefont {C.}~\bibnamefont {Langrenez}}, \bibinfo
  {author} {\bibfnamefont {M.}~\bibnamefont {Lostaglio}}, \bibinfo {author}
  {\bibfnamefont {J.~S.}\ \bibnamefont {Lundeen}}, \ and\ \bibinfo {author}
  {\bibfnamefont {N.~Y.}\ \bibnamefont {Halpern}},\ }\bibinfo {title}
  {Properties and applications of the kirkwood--dirac distribution},\ \href
  {\doibase 10.1088/1367-2630/ada05d} {\bibfield  {journal} {\bibinfo
  {journal} {New Journal of Physics}\ }\textbf {\bibinfo {volume} {26}},\
  \bibinfo {pages} {121201} (\bibinfo {year} {2024})}\BibitemShut {NoStop}%
\bibitem [{\citenamefont {Designolle}\ \emph {et~al.}(2021)\citenamefont
  {Designolle}, \citenamefont {Uola}, \citenamefont {Luoma},\ and\
  \citenamefont {Brunner}}]{Brunner-2021-PRL}%
  \BibitemOpen
  \bibfield  {author} {\bibinfo {author} {\bibfnamefont {S.}~\bibnamefont
  {Designolle}}, \bibinfo {author} {\bibfnamefont {R.}~\bibnamefont {Uola}},
  \bibinfo {author} {\bibfnamefont {K.}~\bibnamefont {Luoma}}, \ and\ \bibinfo
  {author} {\bibfnamefont {N.}~\bibnamefont {Brunner}},\ }\bibinfo {title} {Set
  coherence: Basis-independent quantification of quantum coherence},\ \href
  {\doibase 10.1103/PhysRevLett.126.220404} {\bibfield  {journal} {\bibinfo
  {journal} {Phys. Rev. Lett.}\ }\textbf {\bibinfo {volume} {126}},\ \bibinfo
  {pages} {220404} (\bibinfo {year} {2021})}\BibitemShut {NoStop}%
\bibitem [{\citenamefont {Li}\ \emph {et~al.}(2025)\citenamefont {Li},
  \citenamefont {Wagner},\ and\ \citenamefont {Zhang}}]{li-2025-arxiv}%
  \BibitemOpen
  \bibfield  {author} {\bibinfo {author} {\bibfnamefont {M.-S.}\ \bibnamefont
  {Li}}, \bibinfo {author} {\bibfnamefont {R.}~\bibnamefont {Wagner}}, \ and\
  \bibinfo {author} {\bibfnamefont {L.}~\bibnamefont {Zhang}},\ }\bibinfo
  {title} {Multi-state imaginarity and coherence in qubit systems},\ \href
  {https://arxiv.org/abs/2507.14878} {\bibfield  {journal} {\bibinfo  {journal}
  {arXiv preprint arXiv:2507.14878}\ } (\bibinfo {year} {2025})}\BibitemShut
  {NoStop}%
\bibitem [{\citenamefont {Miyazaki}\ and\ \citenamefont
  {Matsumoto}(2022)}]{Miyazaki-2022-Quantum}%
  \BibitemOpen
  \bibfield  {author} {\bibinfo {author} {\bibfnamefont {J.}~\bibnamefont
  {Miyazaki}}\ and\ \bibinfo {author} {\bibfnamefont {K.}~\bibnamefont
  {Matsumoto}},\ }\bibinfo {title} {Imaginarity-free quantum multiparameter
  estimation},\ \href {\doibase 10.22331/q-2022-03-10-665} {\bibfield
  {journal} {\bibinfo  {journal} {{Quantum}}\ }\textbf {\bibinfo {volume}
  {6}},\ \bibinfo {pages} {665} (\bibinfo {year} {2022})}\BibitemShut {NoStop}%
\bibitem [{\citenamefont {Fernandes}\ \emph {et~al.}(2024)\citenamefont
  {Fernandes}, \citenamefont {Wagner}, \citenamefont {Novo},\ and\
  \citenamefont {Galv\~ao}}]{Fernandes-2024-PRL}%
  \BibitemOpen
  \bibfield  {author} {\bibinfo {author} {\bibfnamefont {C.}~\bibnamefont
  {Fernandes}}, \bibinfo {author} {\bibfnamefont {R.}~\bibnamefont {Wagner}},
  \bibinfo {author} {\bibfnamefont {L.}~\bibnamefont {Novo}}, \ and\ \bibinfo
  {author} {\bibfnamefont {E.~F.}\ \bibnamefont {Galv\~ao}},\ }\bibinfo {title}
  {Unitary-invariant witnesses of quantum imaginarity},\ \href {\doibase
  10.1103/PhysRevLett.133.190201} {\bibfield  {journal} {\bibinfo  {journal}
  {Phys. Rev. Lett.}\ }\textbf {\bibinfo {volume} {133}},\ \bibinfo {pages}
  {190201} (\bibinfo {year} {2024})}\BibitemShut {NoStop}%
\bibitem [{\citenamefont {Wagner}\ and\ \citenamefont
  {Galv\~ao}(2023)}]{Wagner-2023-PRA}%
  \BibitemOpen
  \bibfield  {author} {\bibinfo {author} {\bibfnamefont {R.}~\bibnamefont
  {Wagner}}\ and\ \bibinfo {author} {\bibfnamefont {E.~F.}\ \bibnamefont
  {Galv\~ao}},\ }\bibinfo {title} {Simple proof that anomalous weak values
  require coherence},\ \href {\doibase 10.1103/PhysRevA.108.L040202} {\bibfield
   {journal} {\bibinfo  {journal} {Phys. Rev. A}\ }\textbf {\bibinfo {volume}
  {108}},\ \bibinfo {pages} {L040202} (\bibinfo {year} {2023})}\BibitemShut
  {NoStop}%
\bibitem [{\citenamefont {Yunger~Halpern}\ \emph {et~al.}(2018)\citenamefont
  {Yunger~Halpern}, \citenamefont {Swingle},\ and\ \citenamefont
  {Dressel}}]{Halpern-2018-PRA}%
  \BibitemOpen
  \bibfield  {author} {\bibinfo {author} {\bibfnamefont {N.}~\bibnamefont
  {Yunger~Halpern}}, \bibinfo {author} {\bibfnamefont {B.}~\bibnamefont
  {Swingle}}, \ and\ \bibinfo {author} {\bibfnamefont {J.}~\bibnamefont
  {Dressel}},\ }\bibinfo {title} {Quasiprobability behind the
  out-of-time-ordered correlator},\ \href {\doibase 10.1103/PhysRevA.97.042105}
  {\bibfield  {journal} {\bibinfo  {journal} {Phys. Rev. A}\ }\textbf {\bibinfo
  {volume} {97}},\ \bibinfo {pages} {042105} (\bibinfo {year}
  {2018})}\BibitemShut {NoStop}%
\bibitem [{\citenamefont {Pont}\ \emph {et~al.}(2022)\citenamefont {Pont},
  \citenamefont {Albiero}, \citenamefont {Thomas}, \citenamefont {Spagnolo},
  \citenamefont {Ceccarelli}, \citenamefont {Corrielli}, \citenamefont
  {Brieussel}, \citenamefont {Somaschi}, \citenamefont {Huet}, \citenamefont
  {Harouri}, \citenamefont {Lema\^{\i}tre}, \citenamefont {Sagnes},
  \citenamefont {Belabas}, \citenamefont {Sciarrino}, \citenamefont {Osellame},
  \citenamefont {Senellart},\ and\ \citenamefont {Crespi}}]{Pont-2022-PRX}%
  \BibitemOpen
  \bibfield  {author} {\bibinfo {author} {\bibfnamefont {M.}~\bibnamefont
  {Pont}}, \bibinfo {author} {\bibfnamefont {R.}~\bibnamefont {Albiero}},
  \bibinfo {author} {\bibfnamefont {S.~E.}\ \bibnamefont {Thomas}}, \bibinfo
  {author} {\bibfnamefont {N.}~\bibnamefont {Spagnolo}}, \bibinfo {author}
  {\bibfnamefont {F.}~\bibnamefont {Ceccarelli}}, \bibinfo {author}
  {\bibfnamefont {G.}~\bibnamefont {Corrielli}}, \bibinfo {author}
  {\bibfnamefont {A.}~\bibnamefont {Brieussel}}, \bibinfo {author}
  {\bibfnamefont {N.}~\bibnamefont {Somaschi}}, \bibinfo {author}
  {\bibfnamefont {H.}~\bibnamefont {Huet}}, \bibinfo {author} {\bibfnamefont
  {A.}~\bibnamefont {Harouri}}, \bibinfo {author} {\bibfnamefont
  {A.}~\bibnamefont {Lema\^{\i}tre}}, \bibinfo {author} {\bibfnamefont
  {I.}~\bibnamefont {Sagnes}}, \bibinfo {author} {\bibfnamefont
  {N.}~\bibnamefont {Belabas}}, \bibinfo {author} {\bibfnamefont
  {F.}~\bibnamefont {Sciarrino}}, \bibinfo {author} {\bibfnamefont
  {R.}~\bibnamefont {Osellame}}, \bibinfo {author} {\bibfnamefont
  {P.}~\bibnamefont {Senellart}}, \ and\ \bibinfo {author} {\bibfnamefont
  {A.}~\bibnamefont {Crespi}},\ }\bibinfo {title} {Quantifying $n$-photon
  indistinguishability with a cyclic integrated interferometer},\ \href
  {\doibase 10.1103/PhysRevX.12.031033} {\bibfield  {journal} {\bibinfo
  {journal} {Phys. Rev. X}\ }\textbf {\bibinfo {volume} {12}},\ \bibinfo
  {pages} {031033} (\bibinfo {year} {2022})}\BibitemShut {NoStop}%
\bibitem [{\citenamefont {Oszmaniec}\ \emph {et~al.}(2024)\citenamefont
  {Oszmaniec}, \citenamefont {Brod},\ and\ \citenamefont
  {Galv\~ao}}]{Oszmaniec-2024-NJP}%
  \BibitemOpen
  \bibfield  {author} {\bibinfo {author} {\bibfnamefont {M.}~\bibnamefont
  {Oszmaniec}}, \bibinfo {author} {\bibfnamefont {D.~J.}\ \bibnamefont {Brod}},
  \ and\ \bibinfo {author} {\bibfnamefont {E.~F.}\ \bibnamefont {Galv\~ao}},\
  }\bibinfo {title} {Measuring relational information between quantum states,
  and applications},\ \href {\doibase 10.1088/1367-2630/ad1a27} {\bibfield
  {journal} {\bibinfo  {journal} {New Journal of Physics}\ }\textbf {\bibinfo
  {volume} {26}},\ \bibinfo {pages} {013053} (\bibinfo {year}
  {2024})}\BibitemShut {NoStop}%
\bibitem [{\citenamefont {Wagner}\ \emph {et~al.}(2024)\citenamefont {Wagner},
  \citenamefont {Schwartzman-Nowik}, \citenamefont {Paiva}, \citenamefont
  {Te$'$eni}, \citenamefont {Ruiz-Molero}, \citenamefont {Barbosa},
  \citenamefont {Cohen},\ and\ \citenamefont {Galv{\~a}o}}]{Wagner-2024-QST}%
  \BibitemOpen
  \bibfield  {author} {\bibinfo {author} {\bibfnamefont {R.}~\bibnamefont
  {Wagner}}, \bibinfo {author} {\bibfnamefont {Z.}~\bibnamefont
  {Schwartzman-Nowik}}, \bibinfo {author} {\bibfnamefont {I.~L.}\ \bibnamefont
  {Paiva}}, \bibinfo {author} {\bibfnamefont {A.}~\bibnamefont {Te$'$eni}},
  \bibinfo {author} {\bibfnamefont {A.}~\bibnamefont {Ruiz-Molero}}, \bibinfo
  {author} {\bibfnamefont {R.~S.}\ \bibnamefont {Barbosa}}, \bibinfo {author}
  {\bibfnamefont {E.}~\bibnamefont {Cohen}}, \ and\ \bibinfo {author}
  {\bibfnamefont {E.~F.}\ \bibnamefont {Galv{\~a}o}},\ }\bibinfo {title}
  {Quantum circuits for measuring weak values, kirkwood-dirac quasiprobability
  distributions, and state spectra},\ \href {\doibase 10.1088/2058-9565/ad124c}
  {\bibfield  {journal} {\bibinfo  {journal} {Quantum Science and Technology}\
  }\textbf {\bibinfo {volume} {9}},\ \bibinfo {pages} {015030} (\bibinfo {year}
  {2024})}\BibitemShut {NoStop}%
\bibitem [{\citenamefont {Simonov}\ \emph {et~al.}(2025)\citenamefont
  {Simonov}, \citenamefont {Wagner},\ and\ \citenamefont
  {Galv{\~a}o}}]{Galvao-2025-arxiv}%
  \BibitemOpen
  \bibfield  {author} {\bibinfo {author} {\bibfnamefont {K.}~\bibnamefont
  {Simonov}}, \bibinfo {author} {\bibfnamefont {R.}~\bibnamefont {Wagner}}, \
  and\ \bibinfo {author} {\bibfnamefont {E.}~\bibnamefont {Galv{\~a}o}},\
  }\bibinfo {title} {Estimation of multivariate traces of states given partial
  classical information},\ \href {https://arxiv.org/abs/2505.20208} {\bibfield
  {journal} {\bibinfo  {journal} {arXiv preprint arXiv:2505.20208}\ } (\bibinfo
  {year} {2025})}\BibitemShut {NoStop}%
\bibitem [{\citenamefont {Braunstein}\ and\ \citenamefont {van
  Loock}(2005)}]{Braunstein-2005-RMP}%
  \BibitemOpen
  \bibfield  {author} {\bibinfo {author} {\bibfnamefont {S.~L.}\ \bibnamefont
  {Braunstein}}\ and\ \bibinfo {author} {\bibfnamefont {P.}~\bibnamefont {van
  Loock}},\ }\bibinfo {title} {Quantum information with continuous variables},\
  \href {\doibase 10.1103/RevModPhys.77.513} {\bibfield  {journal} {\bibinfo
  {journal} {Rev. Mod. Phys.}\ }\textbf {\bibinfo {volume} {77}},\ \bibinfo
  {pages} {513} (\bibinfo {year} {2005})}\BibitemShut {NoStop}%
\bibitem [{\citenamefont {Wang}\ \emph {et~al.}(2007)\citenamefont {Wang},
  \citenamefont {Hiroshima}, \citenamefont {Tomita},\ and\ \citenamefont
  {Hayashi}}]{Wang-2007-PR}%
  \BibitemOpen
  \bibfield  {author} {\bibinfo {author} {\bibfnamefont {X.-B.}\ \bibnamefont
  {Wang}}, \bibinfo {author} {\bibfnamefont {T.}~\bibnamefont {Hiroshima}},
  \bibinfo {author} {\bibfnamefont {A.}~\bibnamefont {Tomita}}, \ and\ \bibinfo
  {author} {\bibfnamefont {M.}~\bibnamefont {Hayashi}},\ }\bibinfo {title}
  {Quantum information with gaussian states},\ \href {\doibase
  https://doi.org/10.1016/j.physrep.2007.04.005} {\bibfield  {journal}
  {\bibinfo  {journal} {Physics Reports}\ }\textbf {\bibinfo {volume} {448}},\
  \bibinfo {pages} {1} (\bibinfo {year} {2007})}\BibitemShut {NoStop}%
\bibitem [{\citenamefont {Ferraro}\ \emph {et~al.}(2005)\citenamefont
  {Ferraro}, \citenamefont {Olivares},\ and\ \citenamefont
  {Paris}}]{Ferraro-2005-arxiv}%
  \BibitemOpen
  \bibfield  {author} {\bibinfo {author} {\bibfnamefont {A.}~\bibnamefont
  {Ferraro}}, \bibinfo {author} {\bibfnamefont {S.}~\bibnamefont {Olivares}}, \
  and\ \bibinfo {author} {\bibfnamefont {M.~G.}\ \bibnamefont {Paris}},\
  }\bibinfo {title} {Gaussian states in continuous variable quantum
  information},\ \href {https://arxiv.org/abs/quant-ph/0503237} {\bibfield
  {journal} {\bibinfo  {journal} {arXiv preprint quant-ph/0503237}\ } (\bibinfo
  {year} {2005})}\BibitemShut {NoStop}%
\bibitem [{\citenamefont {Olivares}(2012)}]{Olivares-2012-EPJ}%
  \BibitemOpen
  \bibfield  {author} {\bibinfo {author} {\bibfnamefont {S.}~\bibnamefont
  {Olivares}},\ }\bibinfo {title} {Quantum optics in the phase space: a
  tutorial on gaussian states},\ \href
  {https://link.springer.com/article/10.1140/epjst/e2012-01532-4} {\bibfield
  {journal} {\bibinfo  {journal} {The European Physical Journal Special
  Topics}\ }\textbf {\bibinfo {volume} {203}},\ \bibinfo {pages} {3} (\bibinfo
  {year} {2012})}\BibitemShut {NoStop}%
\bibitem [{\citenamefont {Weedbrook}\ \emph {et~al.}(2012)\citenamefont
  {Weedbrook}, \citenamefont {Pirandola}, \citenamefont {Garc\'{\i}a-Patr\'on},
  \citenamefont {Cerf}, \citenamefont {Ralph}, \citenamefont {Shapiro},\ and\
  \citenamefont {Lloyd}}]{Weedbrook-2012-RMP}%
  \BibitemOpen
  \bibfield  {author} {\bibinfo {author} {\bibfnamefont {C.}~\bibnamefont
  {Weedbrook}}, \bibinfo {author} {\bibfnamefont {S.}~\bibnamefont
  {Pirandola}}, \bibinfo {author} {\bibfnamefont {R.}~\bibnamefont
  {Garc\'{\i}a-Patr\'on}}, \bibinfo {author} {\bibfnamefont {N.~J.}\
  \bibnamefont {Cerf}}, \bibinfo {author} {\bibfnamefont {T.~C.}\ \bibnamefont
  {Ralph}}, \bibinfo {author} {\bibfnamefont {J.~H.}\ \bibnamefont {Shapiro}},
  \ and\ \bibinfo {author} {\bibfnamefont {S.}~\bibnamefont {Lloyd}},\
  }\bibinfo {title} {Gaussian quantum information},\ \href {\doibase
  10.1103/RevModPhys.84.621} {\bibfield  {journal} {\bibinfo  {journal} {Rev.
  Mod. Phys.}\ }\textbf {\bibinfo {volume} {84}},\ \bibinfo {pages} {621}
  (\bibinfo {year} {2012})}\BibitemShut {NoStop}%
\bibitem [{\citenamefont {Adesso}\ \emph {et~al.}(2014)\citenamefont {Adesso},
  \citenamefont {Ragy},\ and\ \citenamefont {Lee}}]{Adesso-2014-OSID}%
  \BibitemOpen
  \bibfield  {author} {\bibinfo {author} {\bibfnamefont {G.}~\bibnamefont
  {Adesso}}, \bibinfo {author} {\bibfnamefont {S.}~\bibnamefont {Ragy}}, \ and\
  \bibinfo {author} {\bibfnamefont {A.~R.}\ \bibnamefont {Lee}},\ }\bibinfo
  {title} {Continuous variable quantum information: Gaussian states and
  beyond},\ \href {\doibase 10.1142/S1230161214400010} {\bibfield  {journal}
  {\bibinfo  {journal} {Open Systems \& Information Dynamics}\ }\textbf
  {\bibinfo {volume} {21}},\ \bibinfo {pages} {1440001} (\bibinfo {year}
  {2014})}\BibitemShut {NoStop}%
\bibitem [{\citenamefont {Serafini}(2023)}]{Serafini-2023-book}%
  \BibitemOpen
  \bibfield  {author} {\bibinfo {author} {\bibfnamefont {A.}~\bibnamefont
  {Serafini}},\ }\href {\doibase 10.1201/9781003250975} {\emph {\bibinfo
  {title} {Quantum continuous variables: a primer of theoretical methods, 2nd
  Edition}}}\ (\bibinfo  {publisher} {CRC press},\ \bibinfo {year}
  {2023})\BibitemShut {NoStop}%
\bibitem [{\citenamefont {Xu}(2016)}]{Xu-2016-PRA}%
  \BibitemOpen
  \bibfield  {author} {\bibinfo {author} {\bibfnamefont {J.}~\bibnamefont
  {Xu}},\ }\bibinfo {title} {Quantifying coherence of gaussian states},\ \href
  {\doibase 10.1103/PhysRevA.93.032111} {\bibfield  {journal} {\bibinfo
  {journal} {Phys. Rev. A}\ }\textbf {\bibinfo {volume} {93}},\ \bibinfo
  {pages} {032111} (\bibinfo {year} {2016})}\BibitemShut {NoStop}%
\bibitem [{\citenamefont {Quesada}\ \emph {et~al.}(2019)\citenamefont
  {Quesada}, \citenamefont {Helt}, \citenamefont {Izaac}, \citenamefont
  {Arrazola}, \citenamefont {Shahrokhshahi}, \citenamefont {Myers},\ and\
  \citenamefont {Sabapathy}}]{Quesada-2019-PRA}%
  \BibitemOpen
  \bibfield  {author} {\bibinfo {author} {\bibfnamefont {N.}~\bibnamefont
  {Quesada}}, \bibinfo {author} {\bibfnamefont {L.~G.}\ \bibnamefont {Helt}},
  \bibinfo {author} {\bibfnamefont {J.}~\bibnamefont {Izaac}}, \bibinfo
  {author} {\bibfnamefont {J.~M.}\ \bibnamefont {Arrazola}}, \bibinfo {author}
  {\bibfnamefont {R.}~\bibnamefont {Shahrokhshahi}}, \bibinfo {author}
  {\bibfnamefont {C.~R.}\ \bibnamefont {Myers}}, \ and\ \bibinfo {author}
  {\bibfnamefont {K.~K.}\ \bibnamefont {Sabapathy}},\ }\bibinfo {title}
  {Simulating realistic non-gaussian state preparation},\ \href {\doibase
  10.1103/PhysRevA.100.022341} {\bibfield  {journal} {\bibinfo  {journal}
  {Phys. Rev. A}\ }\textbf {\bibinfo {volume} {100}},\ \bibinfo {pages}
  {022341} (\bibinfo {year} {2019})}\BibitemShut {NoStop}%
\bibitem [{\citenamefont {Yao}\ \emph {et~al.}(2024)\citenamefont {Yao},
  \citenamefont {Miatto},\ and\ \citenamefont {Quesada}}]{Quesada-2024-SP}%
  \BibitemOpen
  \bibfield  {author} {\bibinfo {author} {\bibfnamefont {Y.}~\bibnamefont
  {Yao}}, \bibinfo {author} {\bibfnamefont {F.}~\bibnamefont {Miatto}}, \ and\
  \bibinfo {author} {\bibfnamefont {N.}~\bibnamefont {Quesada}},\ }\bibinfo
  {title} {{Riemannian optimization of photonic quantum circuits in phase and
  Fock space}},\ \href {\doibase 10.21468/SciPostPhys.17.3.082} {\bibfield
  {journal} {\bibinfo  {journal} {SciPost Phys.}\ }\textbf {\bibinfo {volume}
  {17}},\ \bibinfo {pages} {082} (\bibinfo {year} {2024})}\BibitemShut
  {NoStop}%
\bibitem [{\citenamefont {Simon}(2000)}]{Simon-2000-PRL}%
  \BibitemOpen
  \bibfield  {author} {\bibinfo {author} {\bibfnamefont {R.}~\bibnamefont
  {Simon}},\ }\bibinfo {title} {Peres-horodecki separability criterion for
  continuous variable systems},\ \href {\doibase 10.1103/PhysRevLett.84.2726}
  {\bibfield  {journal} {\bibinfo  {journal} {Phys. Rev. Lett.}\ }\textbf
  {\bibinfo {volume} {84}},\ \bibinfo {pages} {2726} (\bibinfo {year}
  {2000})}\BibitemShut {NoStop}%
\bibitem [{\citenamefont {Duan}\ \emph {et~al.}(2000)\citenamefont {Duan},
  \citenamefont {Giedke}, \citenamefont {Cirac},\ and\ \citenamefont
  {Zoller}}]{Duan-2000-PRL}%
  \BibitemOpen
  \bibfield  {author} {\bibinfo {author} {\bibfnamefont {L.-M.}\ \bibnamefont
  {Duan}}, \bibinfo {author} {\bibfnamefont {G.}~\bibnamefont {Giedke}},
  \bibinfo {author} {\bibfnamefont {J.~I.}\ \bibnamefont {Cirac}}, \ and\
  \bibinfo {author} {\bibfnamefont {P.}~\bibnamefont {Zoller}},\ }\bibinfo
  {title} {Inseparability criterion for continuous variable systems},\ \href
  {\doibase 10.1103/PhysRevLett.84.2722} {\bibfield  {journal} {\bibinfo
  {journal} {Phys. Rev. Lett.}\ }\textbf {\bibinfo {volume} {84}},\ \bibinfo
  {pages} {2722} (\bibinfo {year} {2000})}\BibitemShut {NoStop}%
\bibitem [{\citenamefont {Giedke}\ \emph {et~al.}(2001)\citenamefont {Giedke},
  \citenamefont {Kraus}, \citenamefont {Lewenstein},\ and\ \citenamefont
  {Cirac}}]{Cirac-2001-PRL}%
  \BibitemOpen
  \bibfield  {author} {\bibinfo {author} {\bibfnamefont {G.}~\bibnamefont
  {Giedke}}, \bibinfo {author} {\bibfnamefont {B.}~\bibnamefont {Kraus}},
  \bibinfo {author} {\bibfnamefont {M.}~\bibnamefont {Lewenstein}}, \ and\
  \bibinfo {author} {\bibfnamefont {J.~I.}\ \bibnamefont {Cirac}},\ }\bibinfo
  {title} {Entanglement criteria for all bipartite gaussian states},\ \href
  {\doibase 10.1103/PhysRevLett.87.167904} {\bibfield  {journal} {\bibinfo
  {journal} {Phys. Rev. Lett.}\ }\textbf {\bibinfo {volume} {87}},\ \bibinfo
  {pages} {167904} (\bibinfo {year} {2001})}\BibitemShut {NoStop}%
\bibitem [{\citenamefont {Vidal}\ and\ \citenamefont
  {Werner}(2002)}]{Werner-2002-PRA}%
  \BibitemOpen
  \bibfield  {author} {\bibinfo {author} {\bibfnamefont {G.}~\bibnamefont
  {Vidal}}\ and\ \bibinfo {author} {\bibfnamefont {R.~F.}\ \bibnamefont
  {Werner}},\ }\bibinfo {title} {Computable measure of entanglement},\ \href
  {\doibase 10.1103/PhysRevA.65.032314} {\bibfield  {journal} {\bibinfo
  {journal} {Phys. Rev. A}\ }\textbf {\bibinfo {volume} {65}},\ \bibinfo
  {pages} {032314} (\bibinfo {year} {2002})}\BibitemShut {NoStop}%
\bibitem [{\citenamefont {Adesso}\ \emph {et~al.}(2004)\citenamefont {Adesso},
  \citenamefont {Serafini},\ and\ \citenamefont
  {Illuminati}}]{Adesso-2004-PRA}%
  \BibitemOpen
  \bibfield  {author} {\bibinfo {author} {\bibfnamefont {G.}~\bibnamefont
  {Adesso}}, \bibinfo {author} {\bibfnamefont {A.}~\bibnamefont {Serafini}}, \
  and\ \bibinfo {author} {\bibfnamefont {F.}~\bibnamefont {Illuminati}},\
  }\bibinfo {title} {Extremal entanglement and mixedness in continuous variable
  systems},\ \href {\doibase 10.1103/PhysRevA.70.022318} {\bibfield  {journal}
  {\bibinfo  {journal} {Phys. Rev. A}\ }\textbf {\bibinfo {volume} {70}},\
  \bibinfo {pages} {022318} (\bibinfo {year} {2004})}\BibitemShut {NoStop}%
\bibitem [{\citenamefont {Marian}\ and\ \citenamefont
  {Marian}(2008)}]{Marian-2008-PRA}%
  \BibitemOpen
  \bibfield  {author} {\bibinfo {author} {\bibfnamefont {P.}~\bibnamefont
  {Marian}}\ and\ \bibinfo {author} {\bibfnamefont {T.~A.}\ \bibnamefont
  {Marian}},\ }\bibinfo {title} {Bures distance as a measure of entanglement
  for symmetric two-mode gaussian states},\ \href {\doibase
  10.1103/PhysRevA.77.062319} {\bibfield  {journal} {\bibinfo  {journal} {Phys.
  Rev. A}\ }\textbf {\bibinfo {volume} {77}},\ \bibinfo {pages} {062319}
  (\bibinfo {year} {2008})}\BibitemShut {NoStop}%
\bibitem [{\citenamefont {Chen}\ \emph {et~al.}(2023)\citenamefont {Chen},
  \citenamefont {Miao}, \citenamefont {Yin},\ and\ \citenamefont
  {Yuan}}]{Chen-2023-PRA}%
  \BibitemOpen
  \bibfield  {author} {\bibinfo {author} {\bibfnamefont {X.-y.}\ \bibnamefont
  {Chen}}, \bibinfo {author} {\bibfnamefont {M.}~\bibnamefont {Miao}}, \bibinfo
  {author} {\bibfnamefont {R.}~\bibnamefont {Yin}}, \ and\ \bibinfo {author}
  {\bibfnamefont {J.}~\bibnamefont {Yuan}},\ }\bibinfo {title} {Gaussian
  entanglement witness and refined werner-wolf criterion for continuous
  variables},\ \href {\doibase 10.1103/PhysRevA.107.022410} {\bibfield
  {journal} {\bibinfo  {journal} {Phys. Rev. A}\ }\textbf {\bibinfo {volume}
  {107}},\ \bibinfo {pages} {022410} (\bibinfo {year} {2023})}\BibitemShut
  {NoStop}%
\bibitem [{\citenamefont {Adesso}\ and\ \citenamefont
  {Datta}(2010)}]{Adesso-2010-PRL}%
  \BibitemOpen
  \bibfield  {author} {\bibinfo {author} {\bibfnamefont {G.}~\bibnamefont
  {Adesso}}\ and\ \bibinfo {author} {\bibfnamefont {A.}~\bibnamefont {Datta}},\
  }\bibinfo {title} {Quantum versus classical correlations in gaussian
  states},\ \href {\doibase 10.1103/PhysRevLett.105.030501} {\bibfield
  {journal} {\bibinfo  {journal} {Phys. Rev. Lett.}\ }\textbf {\bibinfo
  {volume} {105}},\ \bibinfo {pages} {030501} (\bibinfo {year}
  {2010})}\BibitemShut {NoStop}%
\bibitem [{\citenamefont {Giorda}\ and\ \citenamefont
  {Paris}(2010)}]{Paris-2010-PRL}%
  \BibitemOpen
  \bibfield  {author} {\bibinfo {author} {\bibfnamefont {P.}~\bibnamefont
  {Giorda}}\ and\ \bibinfo {author} {\bibfnamefont {M.~G.~A.}\ \bibnamefont
  {Paris}},\ }\bibinfo {title} {Gaussian quantum discord},\ \href {\doibase
  10.1103/PhysRevLett.105.020503} {\bibfield  {journal} {\bibinfo  {journal}
  {Phys. Rev. Lett.}\ }\textbf {\bibinfo {volume} {105}},\ \bibinfo {pages}
  {020503} (\bibinfo {year} {2010})}\BibitemShut {NoStop}%
\bibitem [{\citenamefont {Pirandola}\ \emph {et~al.}(2014)\citenamefont
  {Pirandola}, \citenamefont {Spedalieri}, \citenamefont {Braunstein},
  \citenamefont {Cerf},\ and\ \citenamefont {Lloyd}}]{Lloyd-2014-PRL}%
  \BibitemOpen
  \bibfield  {author} {\bibinfo {author} {\bibfnamefont {S.}~\bibnamefont
  {Pirandola}}, \bibinfo {author} {\bibfnamefont {G.}~\bibnamefont
  {Spedalieri}}, \bibinfo {author} {\bibfnamefont {S.~L.}\ \bibnamefont
  {Braunstein}}, \bibinfo {author} {\bibfnamefont {N.~J.}\ \bibnamefont
  {Cerf}}, \ and\ \bibinfo {author} {\bibfnamefont {S.}~\bibnamefont {Lloyd}},\
  }\bibinfo {title} {Optimality of gaussian discord},\ \href {\doibase
  10.1103/PhysRevLett.113.140405} {\bibfield  {journal} {\bibinfo  {journal}
  {Physical Review Letters}\ }\textbf {\bibinfo {volume} {113}},\ \bibinfo
  {pages} {140405} (\bibinfo {year} {2014})}\BibitemShut {NoStop}%
\bibitem [{\citenamefont {Albarelli}\ \emph {et~al.}(2017)\citenamefont
  {Albarelli}, \citenamefont {Genoni},\ and\ \citenamefont
  {Paris}}]{Paris-2017-PRA}%
  \BibitemOpen
  \bibfield  {author} {\bibinfo {author} {\bibfnamefont {F.}~\bibnamefont
  {Albarelli}}, \bibinfo {author} {\bibfnamefont {M.~G.}\ \bibnamefont
  {Genoni}}, \ and\ \bibinfo {author} {\bibfnamefont {M.~G.~A.}\ \bibnamefont
  {Paris}},\ }\bibinfo {title} {Generation of coherence via gaussian
  measurements},\ \href {\doibase 10.1103/PhysRevA.96.012337} {\bibfield
  {journal} {\bibinfo  {journal} {Phys. Rev. A}\ }\textbf {\bibinfo {volume}
  {96}},\ \bibinfo {pages} {012337} (\bibinfo {year} {2017})}\BibitemShut
  {NoStop}%
\bibitem [{\citenamefont {Du}\ and\ \citenamefont {Bai}(2022)}]{Du-2022-PRA}%
  \BibitemOpen
  \bibfield  {author} {\bibinfo {author} {\bibfnamefont {S.}~\bibnamefont
  {Du}}\ and\ \bibinfo {author} {\bibfnamefont {Z.}~\bibnamefont {Bai}},\
  }\bibinfo {title} {Conversion of gaussian states under incoherent gaussian
  operations},\ \href {\doibase 10.1103/PhysRevA.105.022412} {\bibfield
  {journal} {\bibinfo  {journal} {Phys. Rev. A}\ }\textbf {\bibinfo {volume}
  {105}},\ \bibinfo {pages} {022412} (\bibinfo {year} {2022})}\BibitemShut
  {NoStop}%
\bibitem [{\citenamefont {Du}\ and\ \citenamefont {Bai}(2023)}]{Du-2023-PRA}%
  \BibitemOpen
  \bibfield  {author} {\bibinfo {author} {\bibfnamefont {S.}~\bibnamefont
  {Du}}\ and\ \bibinfo {author} {\bibfnamefont {Z.}~\bibnamefont {Bai}},\
  }\bibinfo {title} {Incoherent gaussian equivalence of $m$-mode gaussian
  states},\ \href {\doibase 10.1103/PhysRevA.107.012407} {\bibfield  {journal}
  {\bibinfo  {journal} {Phys. Rev. A}\ }\textbf {\bibinfo {volume} {107}},\
  \bibinfo {pages} {012407} (\bibinfo {year} {2023})}\BibitemShut {NoStop}%
\bibitem [{\citenamefont {Hou}\ \emph {et~al.}(2022)\citenamefont {Hou},
  \citenamefont {Liu},\ and\ \citenamefont {Qi}}]{Qi-2022-PRA}%
  \BibitemOpen
  \bibfield  {author} {\bibinfo {author} {\bibfnamefont {J.}~\bibnamefont
  {Hou}}, \bibinfo {author} {\bibfnamefont {L.}~\bibnamefont {Liu}}, \ and\
  \bibinfo {author} {\bibfnamefont {X.}~\bibnamefont {Qi}},\ }\bibinfo {title}
  {Computable multipartite multimode gaussian quantum correlation measure and
  the monogamy relations for continuous-variable systems},\ \href {\doibase
  10.1103/PhysRevA.105.032429} {\bibfield  {journal} {\bibinfo  {journal}
  {Phys. Rev. A}\ }\textbf {\bibinfo {volume} {105}},\ \bibinfo {pages}
  {032429} (\bibinfo {year} {2022})}\BibitemShut {NoStop}%
\bibitem [{\citenamefont {Marian}\ and\ \citenamefont
  {Marian}(2012)}]{Marian-2012-PRA}%
  \BibitemOpen
  \bibfield  {author} {\bibinfo {author} {\bibfnamefont {P.}~\bibnamefont
  {Marian}}\ and\ \bibinfo {author} {\bibfnamefont {T.~A.}\ \bibnamefont
  {Marian}},\ }\bibinfo {title} {Uhlmann fidelity between two-mode gaussian
  states},\ \href {\doibase 10.1103/PhysRevA.86.022340} {\bibfield  {journal}
  {\bibinfo  {journal} {Phys. Rev. A}\ }\textbf {\bibinfo {volume} {86}},\
  \bibinfo {pages} {022340} (\bibinfo {year} {2012})}\BibitemShut {NoStop}%
\bibitem [{\citenamefont {Banchi}\ \emph {et~al.}(2015)\citenamefont {Banchi},
  \citenamefont {Braunstein},\ and\ \citenamefont
  {Pirandola}}]{Banchi-2015-PRL}%
  \BibitemOpen
  \bibfield  {author} {\bibinfo {author} {\bibfnamefont {L.}~\bibnamefont
  {Banchi}}, \bibinfo {author} {\bibfnamefont {S.~L.}\ \bibnamefont
  {Braunstein}}, \ and\ \bibinfo {author} {\bibfnamefont {S.}~\bibnamefont
  {Pirandola}},\ }\bibinfo {title} {Quantum fidelity for arbitrary gaussian
  states},\ \href {\doibase 10.1103/PhysRevLett.115.260501} {\bibfield
  {journal} {\bibinfo  {journal} {Phys. Rev. Lett.}\ }\textbf {\bibinfo
  {volume} {115}},\ \bibinfo {pages} {260501} (\bibinfo {year}
  {2015})}\BibitemShut {NoStop}%
\bibitem [{\citenamefont {Simon}\ \emph {et~al.}(1994)\citenamefont {Simon},
  \citenamefont {Mukunda},\ and\ \citenamefont {Dutta}}]{Simon-1994-PRA}%
  \BibitemOpen
  \bibfield  {author} {\bibinfo {author} {\bibfnamefont {R.}~\bibnamefont
  {Simon}}, \bibinfo {author} {\bibfnamefont {N.}~\bibnamefont {Mukunda}}, \
  and\ \bibinfo {author} {\bibfnamefont {B.}~\bibnamefont {Dutta}},\ }\bibinfo
  {title} {Quantum-noise matrix for multimode systems: U(n) invariance,
  squeezing, and normal forms},\ \href {\doibase 10.1103/PhysRevA.49.1567}
  {\bibfield  {journal} {\bibinfo  {journal} {Phys. Rev. A}\ }\textbf {\bibinfo
  {volume} {49}},\ \bibinfo {pages} {1567} (\bibinfo {year}
  {1994})}\BibitemShut {NoStop}%
\bibitem [{\citenamefont {Xu}(2023)}]{Xu-2023-PRA}%
  \BibitemOpen
  \bibfield  {author} {\bibinfo {author} {\bibfnamefont {J.}~\bibnamefont
  {Xu}},\ }\bibinfo {title} {Imaginarity of gaussian states},\ \href {\doibase
  10.1103/PhysRevA.108.062203} {\bibfield  {journal} {\bibinfo  {journal}
  {Phys. Rev. A}\ }\textbf {\bibinfo {volume} {108}},\ \bibinfo {pages}
  {062203} (\bibinfo {year} {2023})}\BibitemShut {NoStop}%
\bibitem [{\citenamefont {Li}\ and\ \citenamefont {Tan}(2025)}]{Li-2025-PRA}%
  \BibitemOpen
  \bibfield  {author} {\bibinfo {author} {\bibfnamefont {M.-S.}\ \bibnamefont
  {Li}}\ and\ \bibinfo {author} {\bibfnamefont {Y.-X.}\ \bibnamefont {Tan}},\
  }\bibinfo {title} {Bargmann invariants for quantum imaginarity},\ \href
  {\doibase 10.1103/PhysRevA.111.022409} {\bibfield  {journal} {\bibinfo
  {journal} {Phys. Rev. A}\ }\textbf {\bibinfo {volume} {111}},\ \bibinfo
  {pages} {022409} (\bibinfo {year} {2025})}\BibitemShut {NoStop}%
\bibitem [{\citenamefont {Zhang}\ \emph {et~al.}(2025)\citenamefont {Zhang},
  \citenamefont {Xie},\ and\ \citenamefont {Li}}]{Zhang-2025-PRA}%
  \BibitemOpen
  \bibfield  {author} {\bibinfo {author} {\bibfnamefont {L.}~\bibnamefont
  {Zhang}}, \bibinfo {author} {\bibfnamefont {B.}~\bibnamefont {Xie}}, \ and\
  \bibinfo {author} {\bibfnamefont {B.}~\bibnamefont {Li}},\ }\bibinfo {title}
  {Geometry of sets of bargmann invariants},\ \href {\doibase
  10.1103/PhysRevA.111.042417} {\bibfield  {journal} {\bibinfo  {journal}
  {Phys. Rev. A}\ }\textbf {\bibinfo {volume} {111}},\ \bibinfo {pages}
  {042417} (\bibinfo {year} {2025})}\BibitemShut {NoStop}%
\bibitem [{\citenamefont {Xu}(2025)}]{Xu-2025-arxiv}%
  \BibitemOpen
  \bibfield  {author} {\bibinfo {author} {\bibfnamefont {J.}~\bibnamefont
  {Xu}},\ }\bibinfo {title} {Numerical ranges of bargmann invariants},\ \href
  {https://arxiv.org/abs/2506.13266} {\bibfield  {journal} {\bibinfo  {journal}
  {arXiv preprint arXiv:2506.13266}\ } (\bibinfo {year} {2025})}\BibitemShut
  {NoStop}%
\bibitem [{\citenamefont {Pratapsi}\ \emph {et~al.}(2025)\citenamefont
  {Pratapsi}, \citenamefont {Gouveia}, \citenamefont {Novo},\ and\
  \citenamefont {Galv\~ao}}]{Pratapsi-2025-arXiv}%
  \BibitemOpen
  \bibfield  {author} {\bibinfo {author} {\bibfnamefont {S.~S.}\ \bibnamefont
  {Pratapsi}}, \bibinfo {author} {\bibfnamefont {J.}~\bibnamefont {Gouveia}},
  \bibinfo {author} {\bibfnamefont {L.}~\bibnamefont {Novo}}, \ and\ \bibinfo
  {author} {\bibfnamefont {E.~F.}\ \bibnamefont {Galv\~ao}},\ }\bibinfo {title}
  {An elementary characterization of bargmann invariants},\ \href
  {https://arxiv.org/abs/2506.17132} {\bibfield  {journal} {\bibinfo  {journal}
  {arXiv preprint arXiv:2506.17132}\ } (\bibinfo {year} {2025})}\BibitemShut
  {NoStop}%
\bibitem [{\citenamefont {B\"{u}nger}\ and\ \citenamefont
  {Rump}(2017)}]{Rump-2017-LAA}%
  \BibitemOpen
  \bibfield  {author} {\bibinfo {author} {\bibfnamefont {F.}~\bibnamefont
  {B\"{u}nger}}\ and\ \bibinfo {author} {\bibfnamefont {S.}~\bibnamefont
  {Rump}},\ }\bibinfo {title} {Yet more elementary proofs that the determinant
  of a symplectic matrix is 1},\ \href {\doibase
  https://doi.org/10.1016/j.laa.2016.11.014} {\bibfield  {journal} {\bibinfo
  {journal} {Linear Algebra and its Applications}\ }\textbf {\bibinfo {volume}
  {515}},\ \bibinfo {pages} {87} (\bibinfo {year} {2017})}\BibitemShut
  {NoStop}%
\bibitem [{\citenamefont {Barnett}\ and\ \citenamefont
  {Radmore}(2002)}]{Barnett-2002-book}%
  \BibitemOpen
  \bibfield  {author} {\bibinfo {author} {\bibfnamefont {S.}~\bibnamefont
  {Barnett}}\ and\ \bibinfo {author} {\bibfnamefont {P.~M.}\ \bibnamefont
  {Radmore}},\ }\href@noop {} {\emph {\bibinfo {title} {Methods in theoretical
  quantum optics}}}\ (\bibinfo  {publisher} {Oxford University Press},\
  \bibinfo {year} {2002})\BibitemShut {NoStop}%
\bibitem [{\citenamefont {Gerry}\ and\ \citenamefont
  {Knight}(2004)}]{Knight-2004-book}%
  \BibitemOpen
  \bibfield  {author} {\bibinfo {author} {\bibfnamefont {C.~C.}\ \bibnamefont
  {Gerry}}\ and\ \bibinfo {author} {\bibfnamefont {P.~L.}\ \bibnamefont
  {Knight}},\ }\href@noop {} {\emph {\bibinfo {title} {Introductory quantum
  optics}}}\ (\bibinfo  {publisher} {Cambridge university press},\ \bibinfo
  {year} {2004})\BibitemShut {NoStop}%
\bibitem [{\citenamefont {Folland}(1989)}]{Folland-1989-book}%
  \BibitemOpen
  \bibfield  {author} {\bibinfo {author} {\bibfnamefont {G.~B.}\ \bibnamefont
  {Folland}},\ }\href@noop {} {\emph {\bibinfo {title} {Harmonic analysis in
  phase space}}},\ \bibinfo {number} {pages 256-257, Appendix A, Theorem 1}\
  (\bibinfo  {publisher} {Princeton university press},\ \bibinfo {year}
  {1989})\BibitemShut {NoStop}%
\bibitem [{\citenamefont {Cahill}\ and\ \citenamefont
  {Glauber}(1969)}]{Glauber-1969-PR}%
  \BibitemOpen
  \bibfield  {author} {\bibinfo {author} {\bibfnamefont {K.~E.}\ \bibnamefont
  {Cahill}}\ and\ \bibinfo {author} {\bibfnamefont {R.~J.}\ \bibnamefont
  {Glauber}},\ }\bibinfo {title} {Ordered expansions in boson amplitude
  operators},\ \href {\doibase 10.1103/PhysRev.177.1857} {\bibfield  {journal}
  {\bibinfo  {journal} {Phys. Rev.}\ }\textbf {\bibinfo {volume} {177}},\
  \bibinfo {pages} {1857} (\bibinfo {year} {1969})}\BibitemShut {NoStop}%
\end{thebibliography}

%

\end{document}